\documentclass[useAMS,usenatbib,usegraphicx]{mn2e}

\usepackage{slashbox}

\title[The rotation of CoRoT field stars]{The rotation of field stars from CoRoT\thanks{The CoRoT  space mission was developed and is operated by the
French space agency CNES, with participation of ESA's RSSD and Science Programmes,
Austria, Belgium, Brazil, Germany, and Spain.} data}
\author[L. Affer, G. Micela, F. Favata, E. Flaccomio]{L. Affer$^{1}$\thanks{E-mail:
affer@astropa.inaf.it},  G. Micela$^{1}$, F. Favata$^{2}$, E. Flaccomio$^{1}$\\
$^{1}$Istituto Nazionale di Astrofisica, Osservatorio Astronomico di Palermo G.\,S. Vaiana, Piazza del Parlamento 1, 90134 Palermo, Italy\\
$^{2}$European Space Agency, 8-10 rue Mario Nikis, 75738 Paris Cedex 15, France}
\begin{document}

\date{Accepted 22 February 2012 Received 22 February 2012; in original form 14 July 2011}

\pagerange{\pageref{firstpage}--\pageref{lastpage}} \pubyear{2002}

\maketitle

\label{firstpage}

\begin{abstract}
We present period measurements of a large sample of field stars in the solar neighbourhood, observed by CoRoT in two different directions of the Galaxy. 
The presence of a period  was detected using the Scargle Lomb Normalized Periodogram 
technique and the autocorrelation analysis. The assessment of the results has been performed 
through a consistency verification supported by the folded light curve analysis. The data analysis procedure has discarded a non-negligible fraction of light curves due to instrumental artifacts, however
it has allowed us to identify pulsators and binaries among a large number of field stars. We measure a wide range of  periods, from 0.25 to 100 days, most of which are rotation periods.\\ 
\indent The final catalogue 
includes 1978 periods, with 1727 of them identified as rotational periods, 169 are classified as pulsations and 82 as orbital periods of binary systems.
Our sample suffers from selection biases not easily corrected for, thus we do not use the distribution of rotation periods to derive the age distribution of the main-sequence population. Nevertheless,
using rotation as a proxy for age, we can identify a sample of young stars ($\le$\, 600 Myr), that will constitute a valuable sample, supported by further spectroscopic observations, to study the recent star formation history in the solar neighborhood.
  \end{abstract}


\section{Introduction}

CoRoT is a pioneering space mission that provided a large scale survey of photometric variability of stars 
with a very high photometric precision and long time coverage. CoRoT observed a 
large sample of stars consisting primarily of main sequence stars in the solar neighbourhood. 
CoRoT primary goals are stellar seismology and extrasolar planets search, nevertheless, stellar light curves 
are themselves a rich source of astrophysical information, thanks to their luminosity modulation. 
There are several reasons for which a star can vary its luminosity, and these may be linked to 
starspot modulation due to stellar magnetic 
activity, to low and high amplitude pulsations, to the existence of close companions (binary 
systems), to the periodic crossing of exoplanets in front of the disk of the star, and so on, including 
a complex mixing of several of the formerly mentioned phenomena. Of particular interest is the 
study of the periodic modulation of starlight produced by non-uniformities on the surface 
of a main-sequence star due to manifestation of stellar activity (e.g., spots and plages) which are 
used to determine the stellar rotation period.\\
\indent Stars contract during the pre-main sequence phase, but the mechanisms governing the angular momentum 
evolution during contraction are still not well understood (depending basically on the star-disk 
interaction). They reach the zero age main sequence (ZAMS) with a maximum velocity, and stars of 
the same spectral type are observed to have a spread of rotational
velocities  \citep{bfa97}.\\ 
\indent During the main sequence (MS) phase rotation decreases with age \citep{wil66,kra67}, due to
the loss of angular momentum through magnetized stellar wind \citep{sch62}, and all the F, G and 
K-type stars tend to converge to a single, color-dependent, rotation-age relation
for stars older than $\sim 10^9 $yr corresponding approximately to the age of the
Hyades.\footnote{The described scenario is valid for single stars, while tidally locked binary systems
maintain fast rotation even at old age.} 
Therefore, when a star reaches this stage, stellar rotation can be used in principle to estimate stellar ages. The first author that determined a 
relation between rotation and age has been \citet{sku72} which obtained  $P_{ROT}\propto\, t^{-0.5}$, where $t$ is the stellar age.  
More recently more complex relations,  called gyrochronology  relations, that take into account also the stellar mass, have been proposed  
\citep{bar03, bar07,mh08}.
As consequence single stars with rotation faster than the convergence value cannot be precisely dated but by them we can put an upper limit of 
$10^9$ yr to their age,  while slower stars of a given mass can be dated just from their rotation.

Using CoRoT photometry we are able to reveal luminosity
variation, with a precision down to 0.1 mmag per hour (magnitude between 11 and 16), during continuous observations 
(up to more than 150 days), allowing to measure photometric periods also in relatively quiet stars (for comparison, the luminosity
variations of the Sun range  between  $\sim$0.3 mmag and $\sim$0.07 mmag at maximum and minimum activity, respectively, \citealt{aig04}). 

The search for periodic modulation changes in the light curves of
dwarf stars observed by CoRoT will allow us to measure
stellar rotation periods in a
large sample of field stars, thus enabling an analysis of
the period as function of spectral type (or stellar mass).\\
\indent \citet{bwb+11} recently performed a similar analysis, discussing the photometric variability characteristics in {\it Kepler} target stars, with a higher photometric precision. They preliminarly conclude that a large number of dwarf stars do vary because of rotation and starspots, and that the stars whose variability is periodic but not due to spots (pulsators, eclipsers) are far less numerous and can be recognized with good efficiency, but warning that further analysis is required to distinguish between various sources of variability in more ambiguous cases.

The immediate goal of the present study is the compilation of a catalogue of periods of main-sequence stars in the CoRoT fields. This catalogue 
will be analyzed in order to study the rotation properties of
solar-like stars in the solar neighbourhood. In particular we are interested in identifying the fast rotator stars 
that, as explained above, are the younger population in the solar neighborhood. This sample will be very valuable for follow-up observations aiming at  reconstructing the star formation history close to the Sun.\\ 
The present paper is organized as follows: in Sects. \ref{obs} and \ref{red}, we describe the CoRoT observations and the
reduction of the light curves. In Sect. \ref{analysis} we describe the methods used to derive
periods, in Sect. \ref{signif} we show the identification of rotational periods and the compiled catalogue. In Sects.
\ref{discuss} and \ref{concl}, we discuss and summarize our findings.

\section[]{CoRoT observations}\label{obs}

CoRoT (COnvection ROtation and planetary Transits, \citealt{baglin06}) is
a space mission dedicated to stellar seismology 
and to the search of extrasolar planets. 
Besides the two main scientific programs, several other ``additional programs" 
are made possible by its high photometric performances and its observing 
runs covering up to five months without interruption. The data presented here have been obtained in the context of the program ``An
unbiased study of rotation and stochastic variability and flaring in all CoRoT\, target stars'' (P.I. F. Favata). 

The CoRoT instrument consists of a telescope with an 
aperture of 27 cm. 
The field of view is a square of 2.8${^\circ}\times 2.8{^\circ}$ 
covered by four 2048 $\times$\, 2048 pixel EEV CCDs, two specialized for the seismology 
mission, the other two for the exoplanet mission.
CoRoT is restricted to point within two 10 degree radius circles, which were selected at the 
intersections of the ecliptic and Galactic planes, one towards the 
Galactic anti-center (RA$_{2000}$=$06{^h}50{^m}25{^s}$, Dec$_{2000}$=$-01{^\circ}42{'}00{''}$) and the other towards 
the Galactic center (RA$_{2000}$=$19{^h}23{^m}34{^s}$, Dec$_{2000}$=$00{^\circ}27{'}36{''}$). 

The exoplanet CCDs are read every 32 s, and photometry for up to 6000 stars per CCD (12\,000 per run) in the range 11.5~$<$ R $<$~16 is performed on-board. 
Light curves with a sampling of 512 s are recorded (32 s sampling is available for up to 500 stars per CCD). For almost all stars brighter than $R$\,=15, 
the flux is separated into broad-band red, green and 
blue channels, containing $\sim$ 40, 30 and 30\% of the flux, respectively (the separation between the red, green and blue bands is adapted for each target stars).

\begin{figure*}
   \includegraphics[width=8cm,height=5cm]{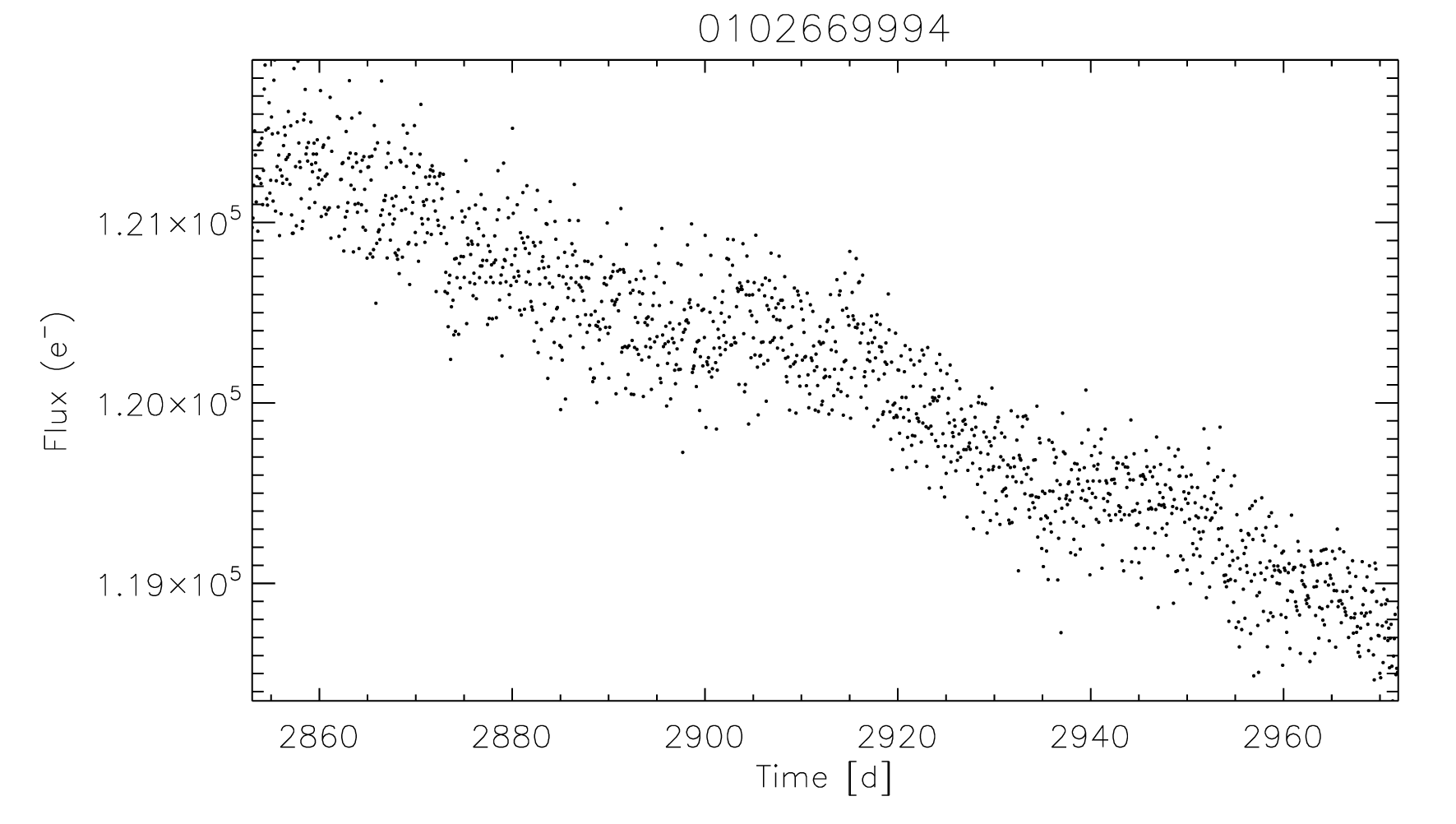}
   \includegraphics[width=8cm,height=5cm]{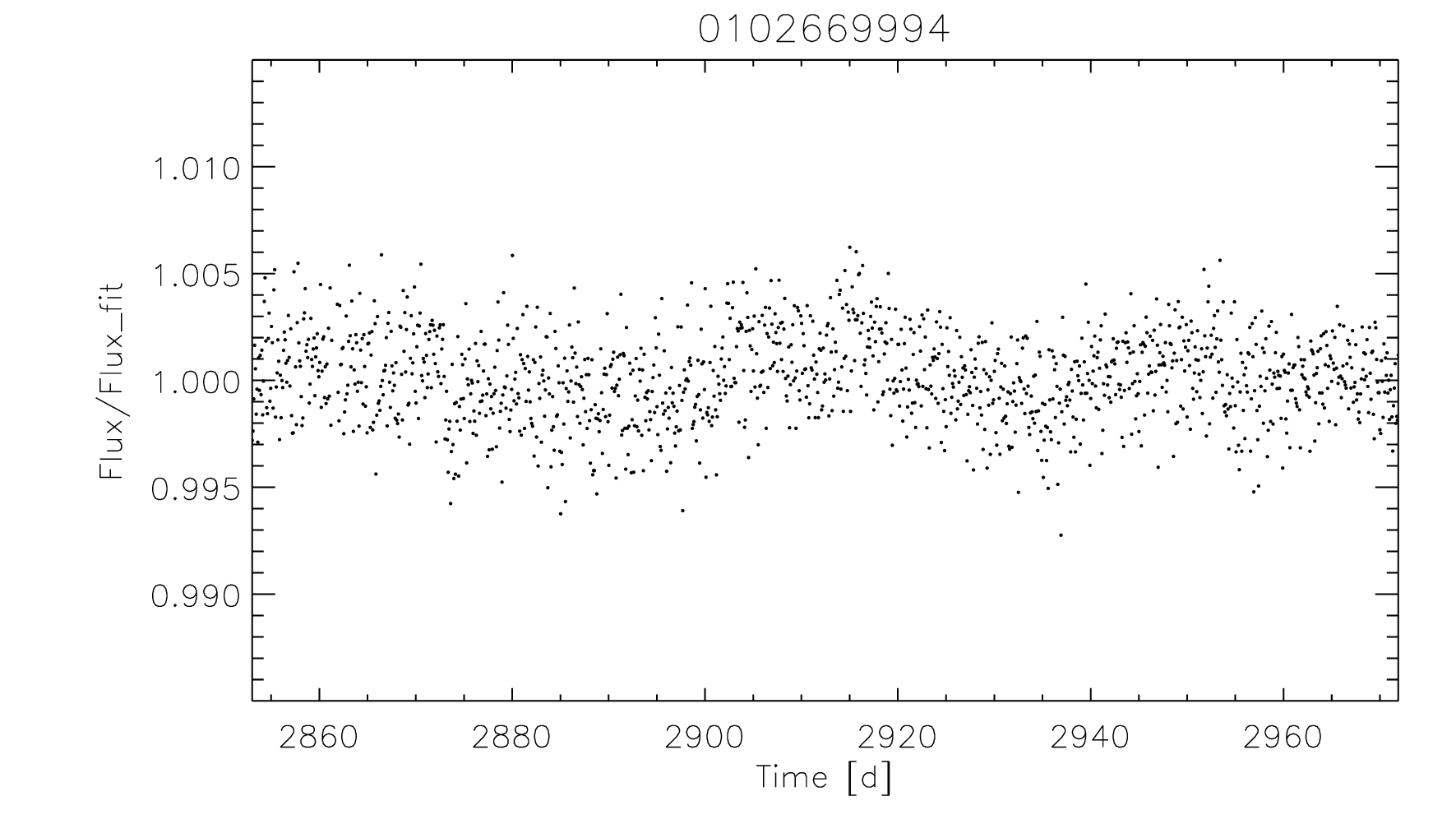}
      \caption{ Example of a raw light curve for a quiet star from the first CoRoT run towards the Galactic anti-center, with a binning of 2h, showing an instrumental long term variation in the $\it{left\,
      panel}$, detrended and normalized in the $\it{right\, panel}$. Time is in days from 2000 January 1.}
	 \label{fig1}
   \end{figure*}

\begin{figure*}
   \includegraphics[width=8cm,height=5cm]{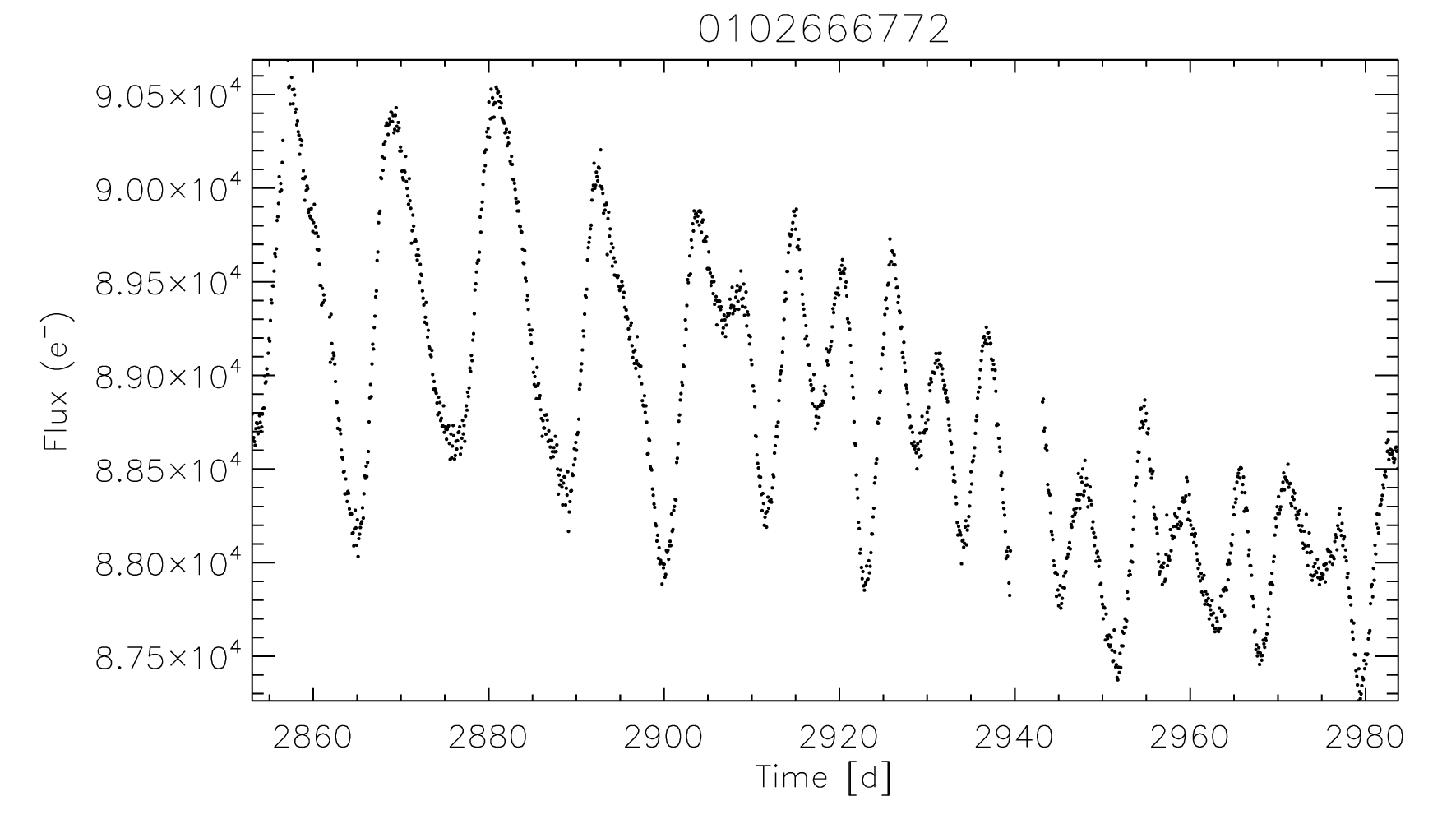}
   \includegraphics[width=8cm,height=5cm]{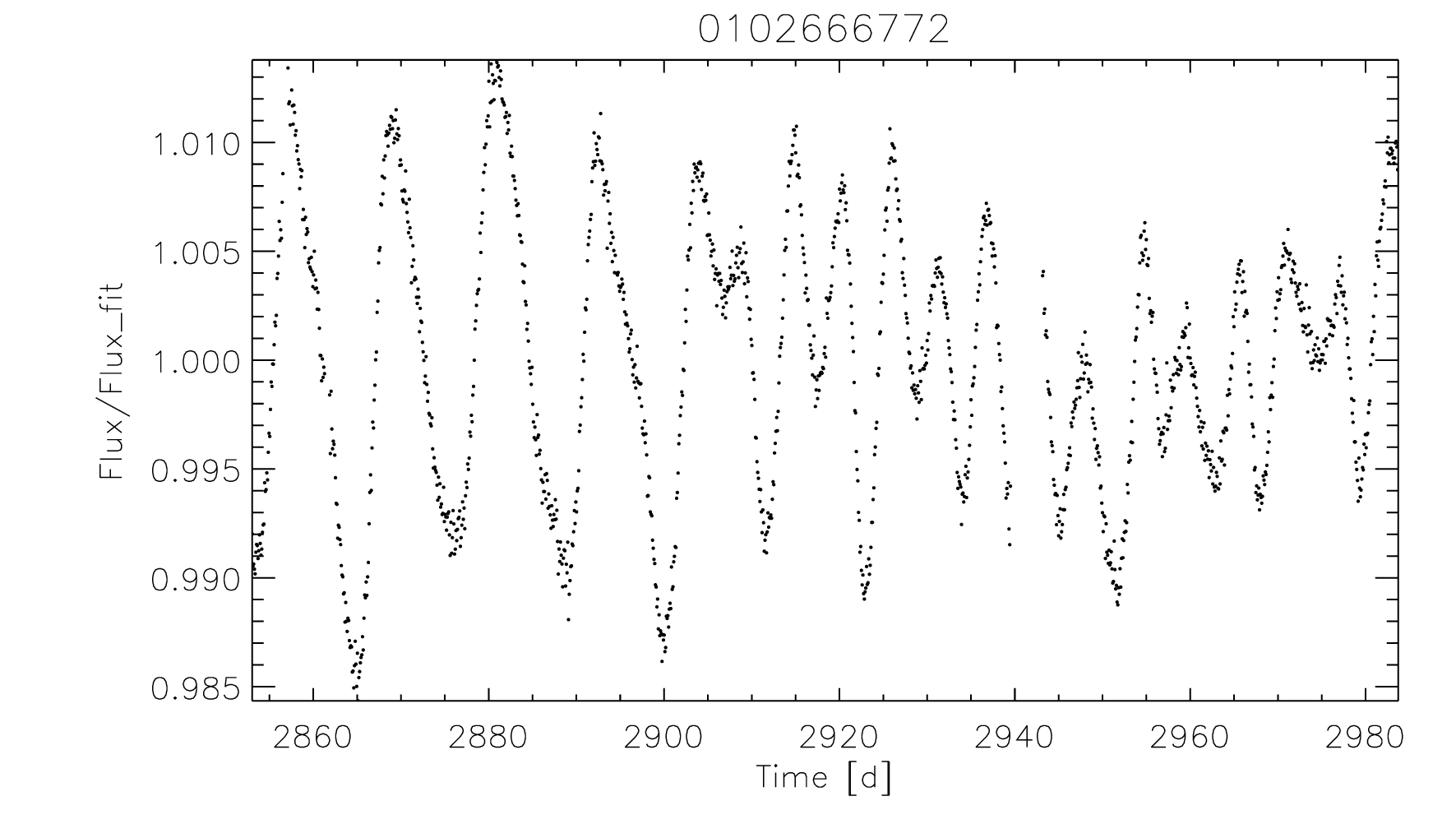}
      \caption{ Same as Fig.~\ref{fig1}, but for an active star.}
	\label{fig2}
   \end{figure*}

The raw data, as downloaded from the satellite (labelled N0 data) are further 
processed on the ground, with the CoRoT pipeline which corrects several effects. Amongst 
these are the electronic offset, gain, electromagnetic interference, and
outliers. The pipeline includes background subtraction and partial 
jitter correction  for the three colour channels,  and also flags data points collected during 
the South Atlantic Anomaly (SAA) crossing. The results of these corrections are labelled N1 data \citep{sfc+07}. 
Further corrections of N1 data, including merging of 512 s and 32 s samplings, 
calculation of heliocentric date, hot pixels detection (data are flagged) lead 
to the N2 data \citep{sfc+07}. The N2 data are the science grade data on which all scientific 
analysis is performed. The light curves are normally sampled at a rate 
of 512 s or oversampled at 32 s; in several cases (in particular for chromatic 
light curves) the oversampling mode was triggered after the start of the run.

In the present paper we analyzed Corot's
first long run observations of the exoplanet field. The first long run 
observations (LRc01) started with a pointing direction close to the center of 
the Galaxy and lasted from May 16 to October 15, 2007 (155 days). 
The first long run observations with a pointing direction close to the anti-center (LRa01) of the 
Galaxy lasted from October 24, 2007 to March 3, 2008 (130 days).
The light curves obtained are 
nearly continuous with only 
a small number of gaps resulting from the periodic crossing of the SAA.\\
\indent In Figures~\ref{fig1} and ~\ref{fig2} we show two examples of N2 data, for quiet and active stars  
with a binning of two hours. The long period trend which is evident in the two examples shown (and in almost all raw light curves) has been removed as described later in Sec.~\ref{red}.

The parent CoRoT sample was selected as a magnitude-limited sample with objects affected by crowding removed. 
The targets observed by CoRoT were selected using the information gathered in the EXODAT database \citep{del06,del09,mdm+07}, 
built with dedicated ground based photometric observations in the visible. Due to the requirements of the exoplanet search it was important to select the 
dwarf population among all potential target stars, in order to increase the chance of detecting planetary transits. Therefore the resulting parent population is 
biased toward dwarf stars even if a substantial fraction of giants may still be present \citep{aig09}. An additional, intrinsic bias present in 
magnitude-limited samples is the increased fraction of older (more luminous and therefore more evolved) stars and binaries (again more luminous than single stars of 
the same age). The effects of these biases on our results will be discussed in Sec.~\ref{discuss}.

\begin{figure*}
   \centering
   \includegraphics[width=7cm]{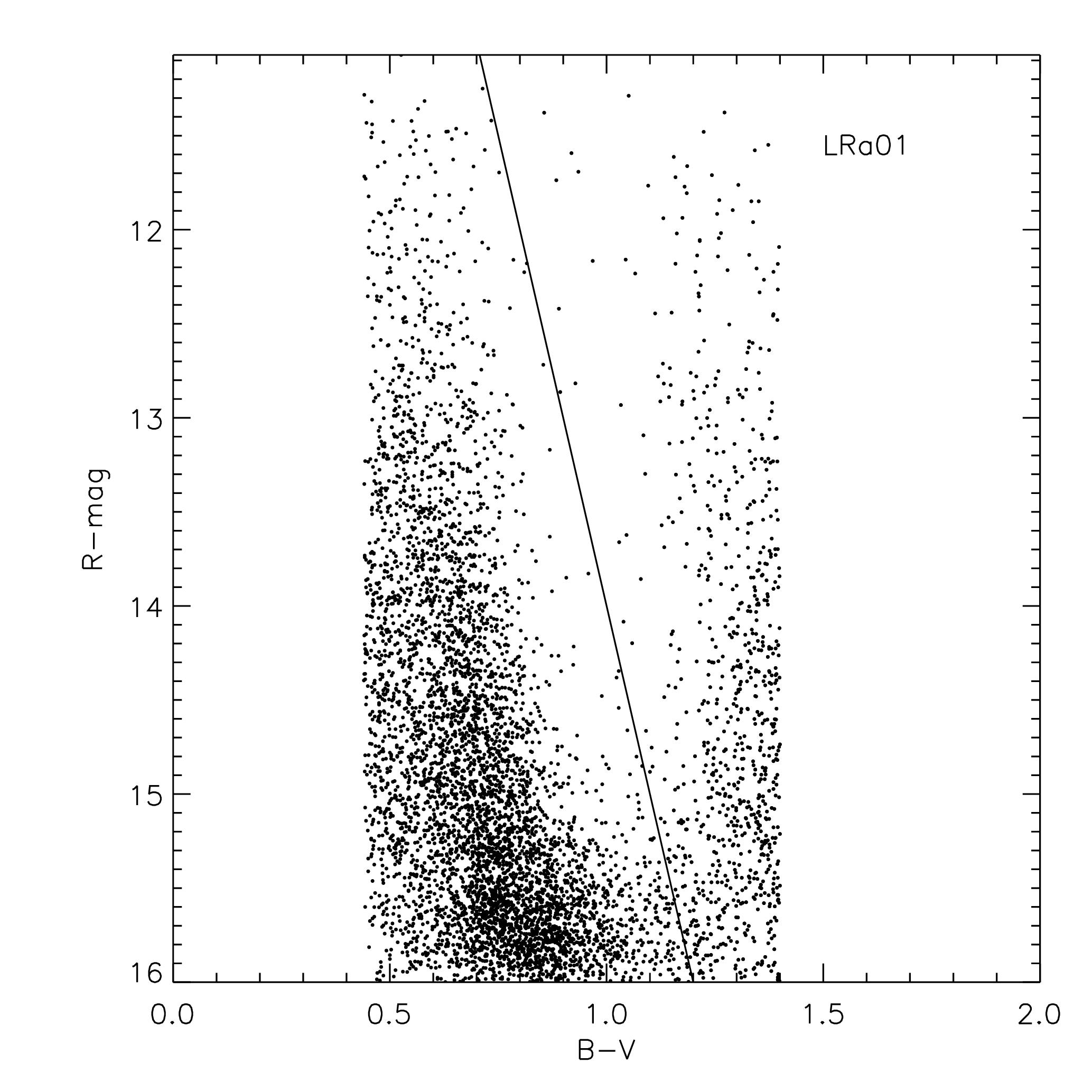}
   \includegraphics[width=7cm]{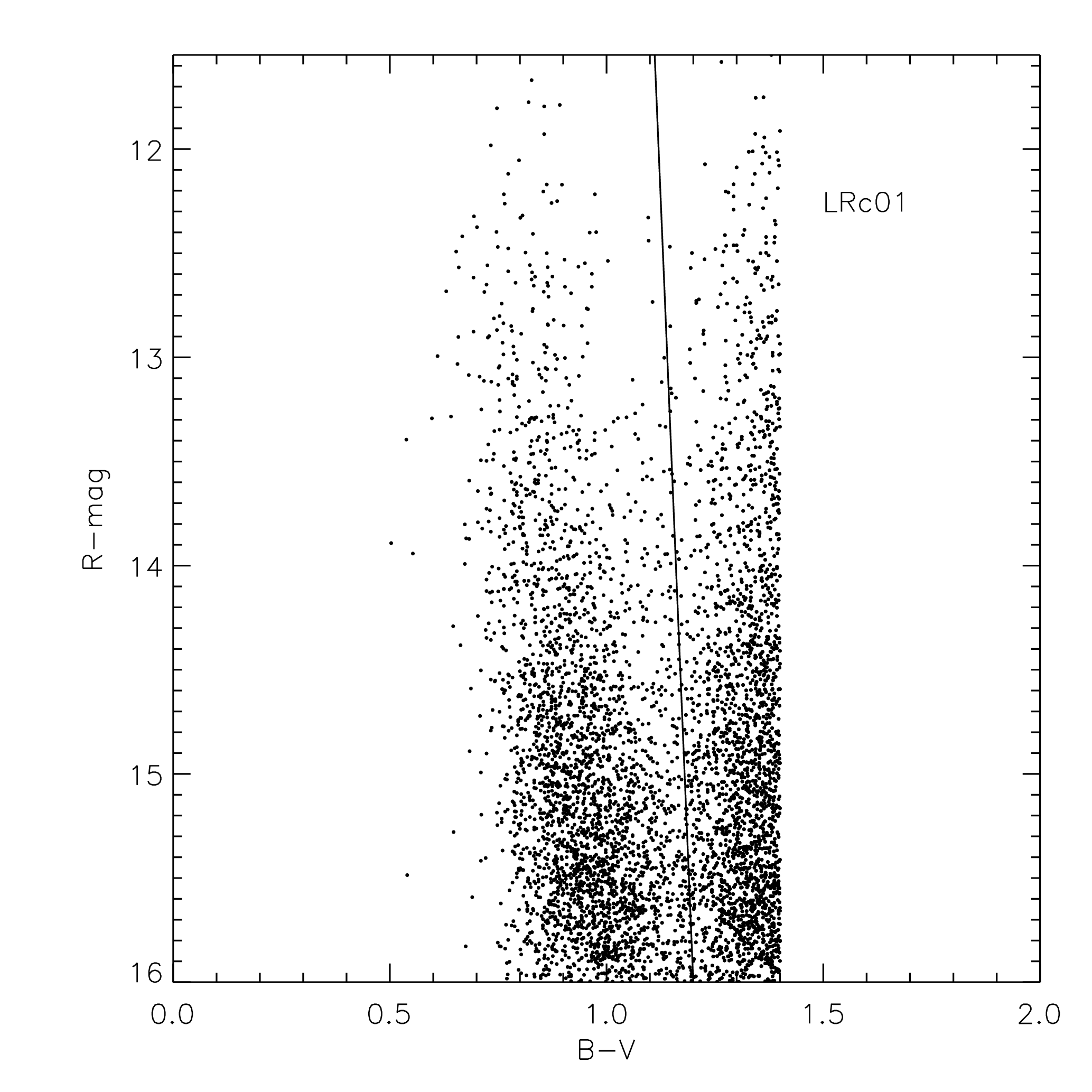}
      \caption{$R$-magnitude versus $B-V$ colour of the selected CoRoT targets in the anti-center ($\it{left}$) and center fields ($\it{right}$). Dots
      on the left of the continuous line correspond to
      stars identified as likely dwarfs, dots on the right of the line to likely giants. The cut between dwarfs and giants was empirically set as a straight line running from $B-V$ = 0.7
      and 1.1 at $R$ = 11 to $B-V$ = 1.2 and 1.2 at $R$ =16 for LRa01 and LRc01, respectively. The larger fraction of giants in LRc01 is clearly visible.
              }
         \label{fig3}
   \end{figure*}

\begin{table}
\centering
\caption{Number of LRa01 ($\it{top}$) and LRc01 ($\it{bottom}$) light curves for main-sequence stars, selected following \citet{aig09}, amongst the different colour indices and magnitude bins.}
\label{tab00}
\begin{tabular}{|c|ccccc|c|} \hline
\backslashbox{B-V}{R}	  &	11-12&	12-13& 13-14& 14-15&15-16& Total\\\hline
0.44-0.58 &	37  &  105  &  303 &   367 &  293&  1105 \\
({\tiny F5--G0}) &	    &    &    &    &   &\\
0.58-0.68 &	18   &	65  &  229 &   423 &  394&   1129 \\
({\tiny G0--G5}) &    &   &   &    &   &\\
0.68-0.81 &	6  &	32  &  121 &   505 &  1042&  1706 \\
({\tiny G5--K0}) &	    &     &    &     &   &\\
0.81-1.15 &	6   &	15  &  17 &   113 &  1191&  1342 \\
({\tiny K0--K5}) &	    &   &    &   &   &\\
1.15-1.40 &	-   &  -  &  - &   10 &  47&  57 \\
({\tiny K5--M0}) &    &    &    &    &   &\\\hline
Total LRa01	  &	67  &  217  & 670 &  1418 & 2967&   5339\\\hline
0.44-0.58 &	-  &  -  &  3 &   - &  1&  4 \\
({\tiny F5--G0}) &	    &     &   &     &   &\\
0.58-0.68 &	-   &	6  &  6 &   2 &  2&   16 \\
({\tiny G0--G5}) &	    &     &   &     &   &\\
0.68-0.81 &	2  &	32  &  86 &   112 &  81&  313 \\
({\tiny G5--K0}) &	    &    &    &    &   &\\
0.81-1.15 &	5   &	48  &  223 &   778 &  1498&  2552 \\
({\tiny K0--K5}) &	    &     &  &     &   &\\
1.15-1.40 &	-   &  -  &  2 &   18 &  97&  117 \\
({\tiny K5--M0}) &	    &   &   &    &   &\\\hline
Total LRc01&	7  &  86  & 320 &  910 & 1679&   3002\\\hline
\end{tabular}
\begin{flushleft}
\end{flushleft}
\end{table}

\section[]{Sample and data reduction}\label{red}

We initially selected from the CoRoT database (Archive of the COROT Data Center at Space
Astrophysics Institute, Orsay-France, http://idoc-corot.ias.u-psud.fr/), observations of dwarf stars (labelled as luminosity class V) with colour index $B-V$ between 0.44 and 1.4 
performed during the LRa01 and LRc01 runs of the exoplanet field (the luminosity class and colour information are from both the EXODAT/COROTSKY database and the archive 
of the CoRoT Data Center). \\
\indent  A significant number of stars with $R< 16$ is expected to be giants and supergiants according to simulations \citep{rrd+03} of the stellar populations in the
field of view. As fully explained in \citet{del09}, \citet{cfo+09}, \citet{aig09} and \citet{cgc+11}, dwarfs and giants are separated in colours, using SED analysis techniques (see \citealt{del09} and references therein, for further
information), 
in color-magnitude or colour-colour diagrams, combining the information of the EXO-DAT
optical photometry  with the near-infrared Two Micron All Sky Survey (2MASS) Point Source catalog \citep{csv+03}. Comparing the results with simulations obtained with a model of synthetic
stellar populations of the Galaxy (the Besancon model, \citealt{rc86,rrd+03}), taking also into account the inhomogeneities of the galactic plane and the mean reddening of the fields \citep{rec+97}, 
they found that the two populations are quite well separated at the brighter end of the CoRoT magnitude range. For a
field towards the center direction which is densely crowded, they estimated that less that 15\% of the stars brighter than $R=$\,14.5 are in the overlapping region between the two groups. There is some
degeneracy for low-luminosity objects as the two populations mix and the separation becomes less reliable, particularly in the center direction. The dividing line between dwarfs and 
giants identified by \citet{aig09} in the $R$ versus $B-V$ colour diagram, as shown in Figure~\ref{fig3}, was obtained from detailed checks with the predictions of galactic models and constitutes an acceptable compromise, from a
statistical point of view, between the loss of some (redder) 
bona fide dwarfs from the sample and the minimization of the residual giant contamination. 
To remove most of the giants from our sample, we eliminated all the stars falling, in the $R$ versus $B-V$ 
colour diagram, to the right of this dividing line.\\
\indent Here we present the rotational analysis of the complete sample of 8341 white and
chromatic (red, blue and green colours) light curves selected with the procedure described above (we eliminated 2934 LCs, classifying them as giants). 
For the aims of our work we did not take into account the colour information and we simply used the white light curve obtained by summing up the fluxes in the three RGB bands.
In particular we analyzed 5339 LRa01 light curves, and 3002 LRc01 light curves (Table~\ref{tab00}).  

We started from the N2 light curves in which most, but not all, instrumental effects have been corrected. The residual artifacts
are:
\begin{enumerate}
\item{long term drifts. These variations are partly due to pointing and to instrumental drift;}
\item{residual periodic variations on the timescale of the satellite orbital period (1.7 h);}
\item{discontinuities due to hot pixels, which often become unstable,
causing further jumps through the run after the first jump (although the pipeline flags almost all of them and 
removes most).}
\end{enumerate}

\begin{figure*}
   \centering
   \includegraphics[width=8cm,height=5cm]{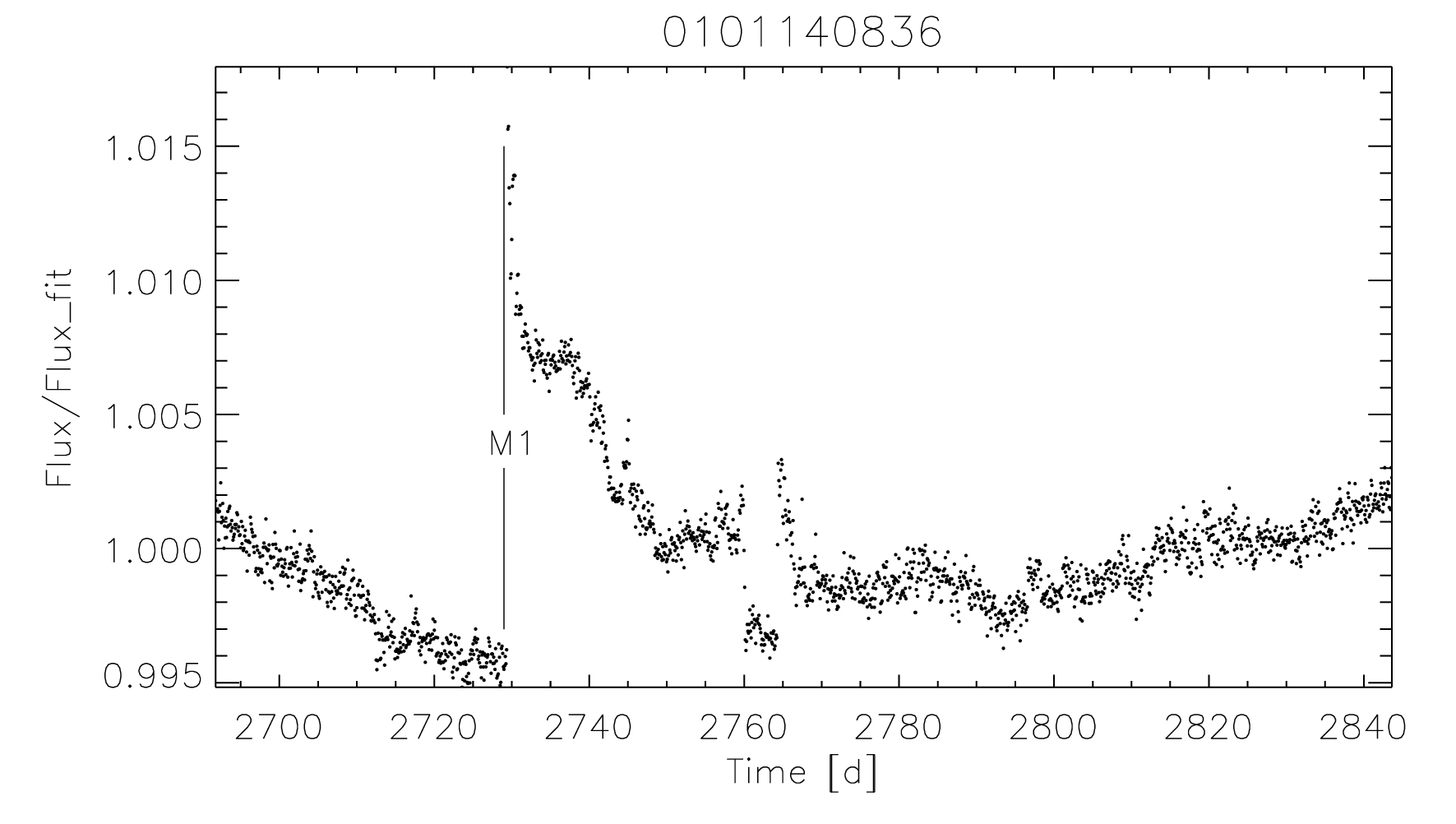}
   \includegraphics[width=8cm,height=5cm]{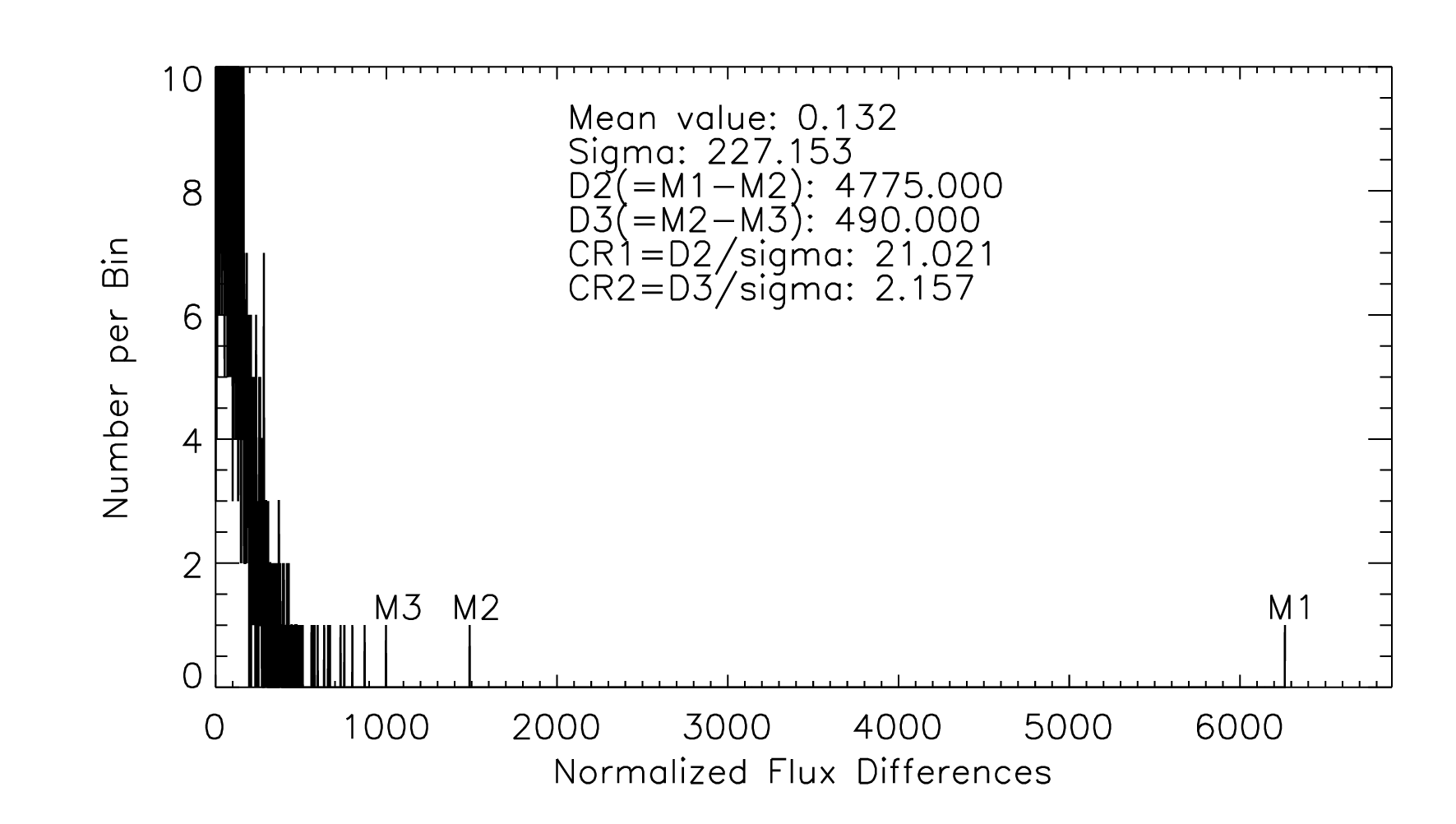}
   \includegraphics[width=8cm,height=5cm]{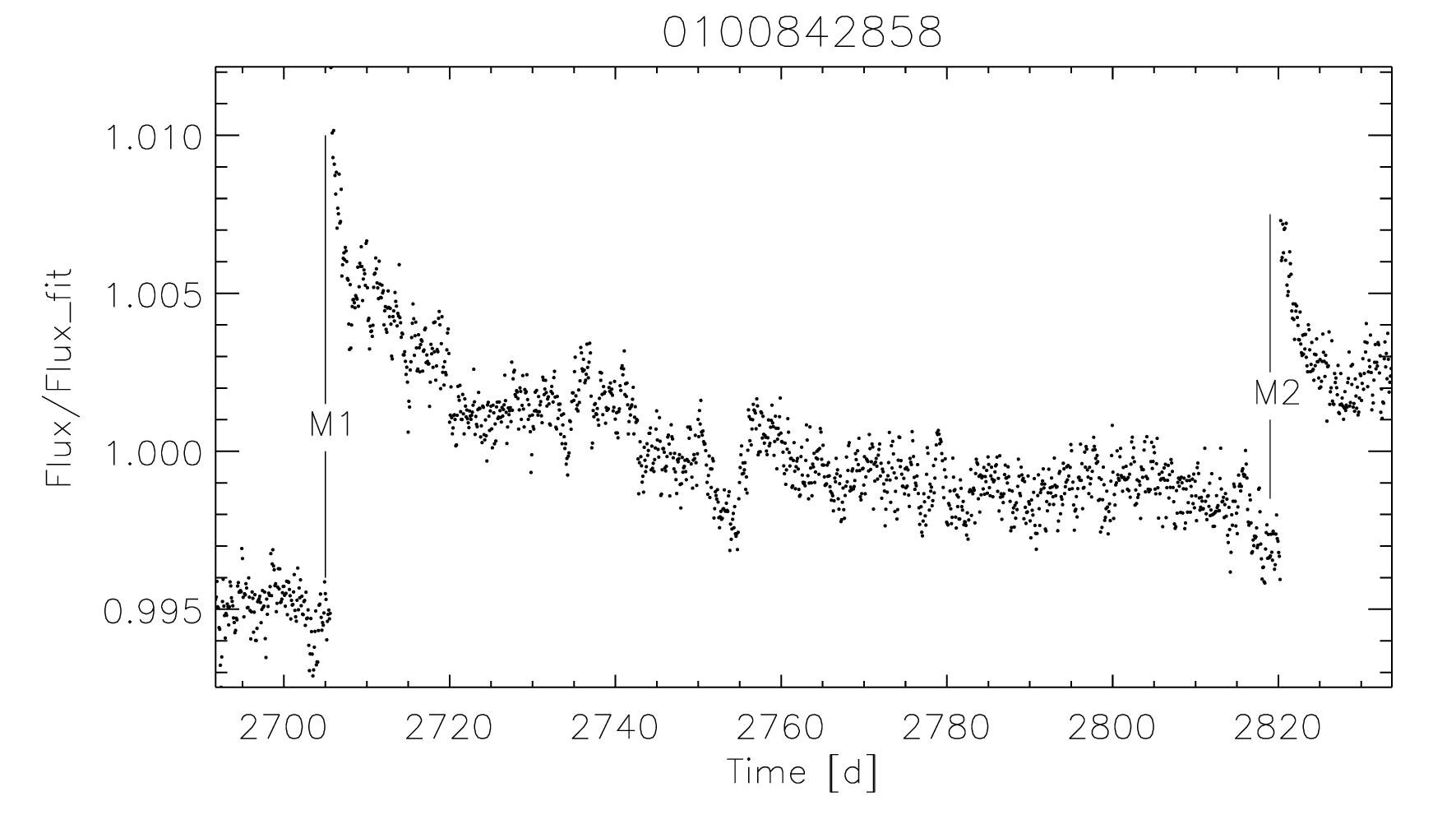}
   \includegraphics[width=8cm,height=5cm]{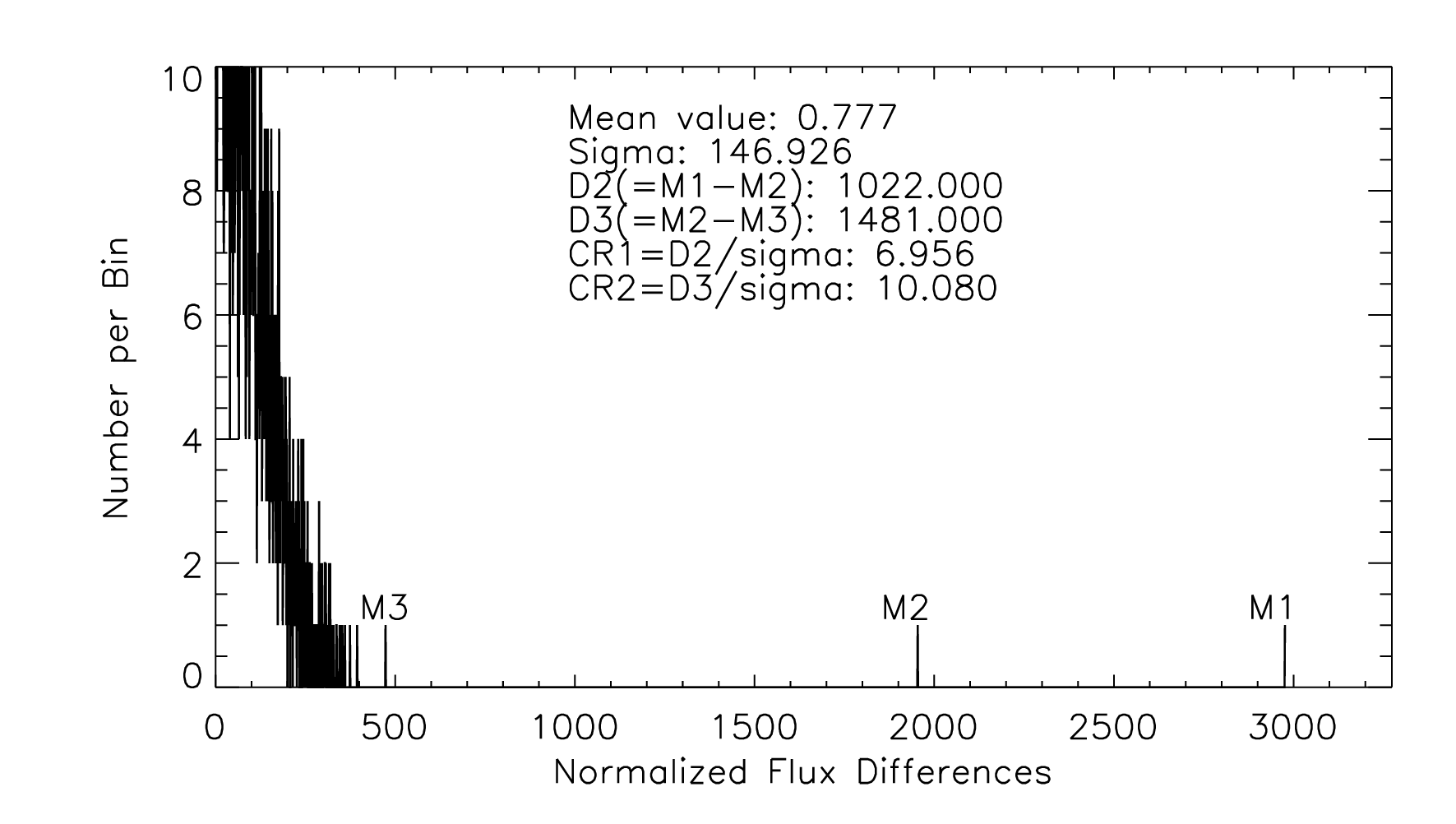}
   \caption{
   $\it{Top\, left\, panel}$: Example of a light curve affected by a hot
pixel event a few days after the start of the
     observations, causing a sudden rise in the measured flux followed by
a gradual decay. Several ``spikes'' (t$\sim $ 2730 d, t$\sim $ 2760
d and t$\sim $ 2765 d) are present in the flux,
     due to high energy particle impact onto the CoRoT CCD. $\it{Top\,
right\, panel}$:
     Distribution of the differences between adjacent fluxes: we
indicated the mean value, the $\sigma_{D}$, the
     differences ($\it{D2,D3}$) between the most deviant values and the two
     rejection criteria $\it{CR_1}$ and $\it{CR_2}$. The criterion used
for rejection is
     $\it{CR_1}$ $>$ 10. $\it{Bottom\, left\, panel}$: In this second
example we underline the presence of two large jumps in
     the flux, at t$\sim $ 2705 d and t$\sim $ 2820. $\it{Bottom\, right\,
     panel}$: The criterion used for the rejection of this light curve is
$\it{CR_2}$ $>$ 10.
}
         \label{fig4}
   \end{figure*}

In order to mitigate the negative effects of these artifacts for our analysis, we preprocessed the light curves in an iterative way. 
We implemented a routine which, as a first step, automatically analyzed all the LCs, detrending, rebinning to 2h and
separating them in two groups according to the presence of discontinuities. As a second step, we visually inspected the results and intervened case-by-case, when necessary, for further correction and selection. In particular: 
\begin{enumerate}
\item{To eliminate the long term drifts we 
detrended the light curves by fitting a third degree polynomial to the data (which works well for most of our LCs) and then dividing the original data points by this
fitting. An example of the result can 
be seen in Figures~\ref{fig1} and ~\ref{fig2}, in which on the left we have the original light curve and the detrended one on the right. To ensure that no strange
behaviour resulted from the process (in particular for the LCs with jumps) the results were inspected by eye, and eventually corrected for the residual instrumental variations with the appropriate fit decided on a case-by-case basis. In particular, for the LCs
with jumps which we judged usable (cutting away the segments with discontinuities, for example) the detrending process was reiterated.}
\item{We rebinned the data to two hours to smooth out the orbital period (1.7 h). Eliminating the signature of the orbital
periodicity reduces the effect of the outliers (mainly due to the satellite crossing of the South Atlantic anomaly).}
\item{ Many CoRoT LCs show sudden large flux discontinuities, and in a discrete number of cases each LC is affected by several ``hot-pixel'' events. If only one pixel is affected, this can be identified by comparing the flux in different
color channels (when available). Different CoRoT users use different techniques to deal with the jump problem \citep{aa08, bfo+07, cf08, rre+08}, and there are several examples of employed strategies also for the ``correction'' of {\it Kepler} LCs with discontinuities
\citep{bwb+10}. Nevertheless, since there is no simple way to correct
all the CoRoT LCs displaying jumps (due to the unpredictable gradual decay which follows each jump and the complicate behaviour of each LC affected by several jumps) without altering the astrophysical signal, we simply
decided to retain only LCs with discontinuities smaller than a selected
threshold, verifying by visual inspection that discontinuities do not affect detrending procedure.
In order to recognize the jumps we have 
first computed the distribution of the differences between fluxes in adjacent bins and evaluated mean value and standard deviation, $\sigma_{D}$. In Figure~\ref{fig4} we report a couple of examples. Since most of the LCs contain more
than one significative jump, we utilize the 3 largest adjacent flux differences
(which we call $\it{M_1}$, $\it{M_2}$ and $\it{M_3}$, see Figure~\ref{fig4}),
and define the parameters for identifying the two discontinuities as follows: $\it{CR_J}=(\it{M_J}-\it{M_{J+1}})/{\sigma_{D}}$, with $\it{J}$=1,2.\\
\indent We selected all the light curves which have both $\it{CR_1}$ and
$\it{CR_2}$ lower than a selected threshold, which we set arbitrarily to
10, which means that we reject light curves with jumps that vary from the mean value by more than 10
standard deviations. There are several LCs, which did not fulfill our criteria for their selection while showing a clear periodicity by visual inspection, and we decided not to reject these LCs. 
Our selection criterion results in about 28\% light curves rejected for a total of 6001 LCs analyzed. The group of rejected LCs were inspected by eye, to evaluate if it was possible to use a shorter segment of the light curve, eliminating that including
the discontinuity, to find a periodicity. We retrieved 240 ``cutted'' LCs, which have been properly corrected for long period trend and have been analyzed serching a periodicity, as well as the rest of the sample.
In Figure~\ref{fig4} two examples of rejected light curves are shown. Top panels show a case
with $\it{CR_1}$ $>$ 10, and bottom panels a case with $\it{CR_2}$ $>$ 10 and $\it{CR_1}$ $<$ 10.} 
\end{enumerate}

Using lower thresholds, 
namely 5 and 3 standard deviations, we would reject about 47\% light curves and about 59\% light 
curves, respectively, but we have verified that final results do not change.
We conservatively decided to adopt 10 $\sigma$\, as threshold. Pathological cases with several 
small jumps and no periodicity are rejected by the autocorrelation analysis (see Sec. \ref{corr}).

\begin{figure*}
   \centering
   \hspace{0.1cm}
   \includegraphics[width=8.cm,height=5cm]{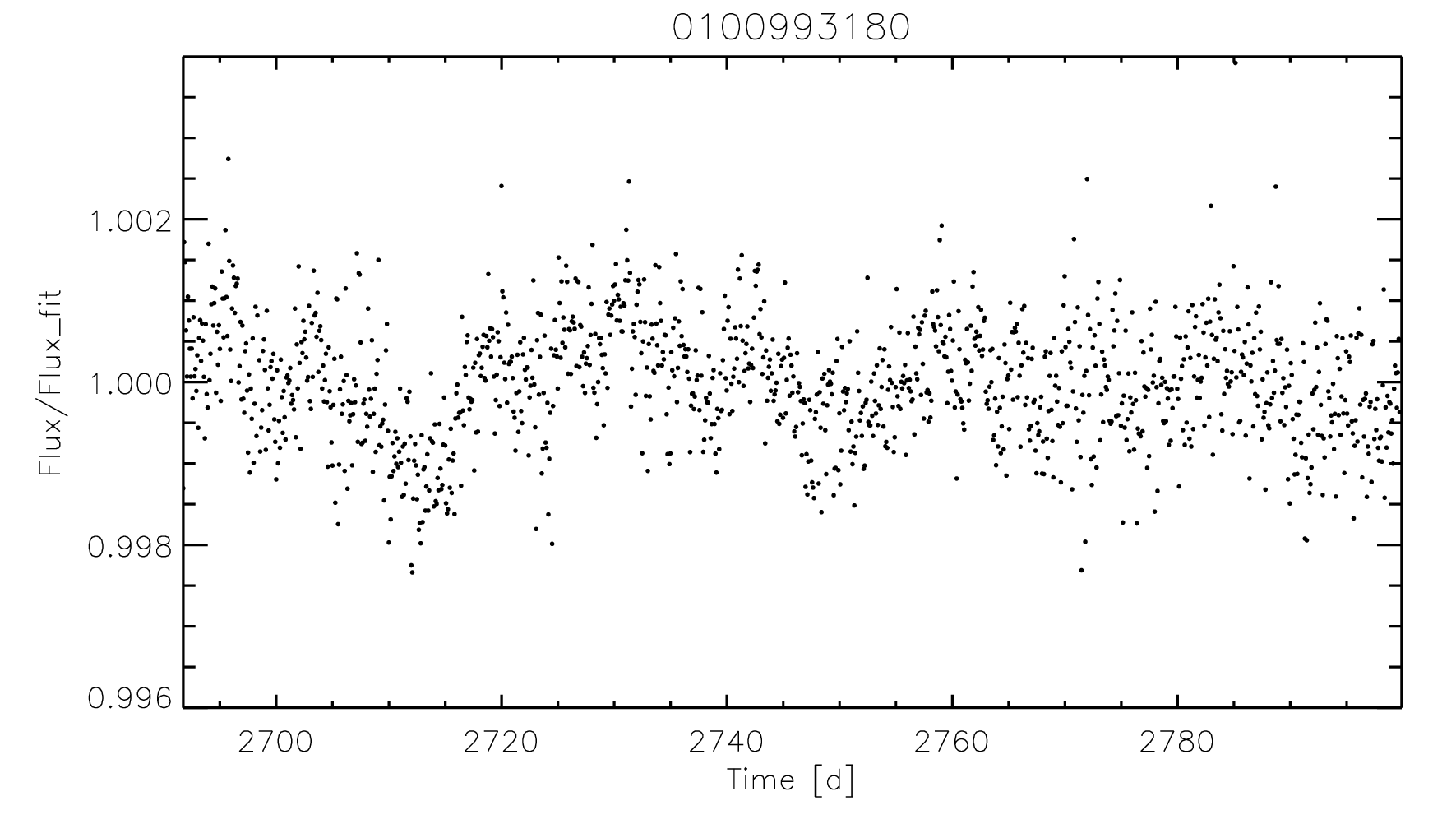}
   \includegraphics[width=8.cm,height=5cm]{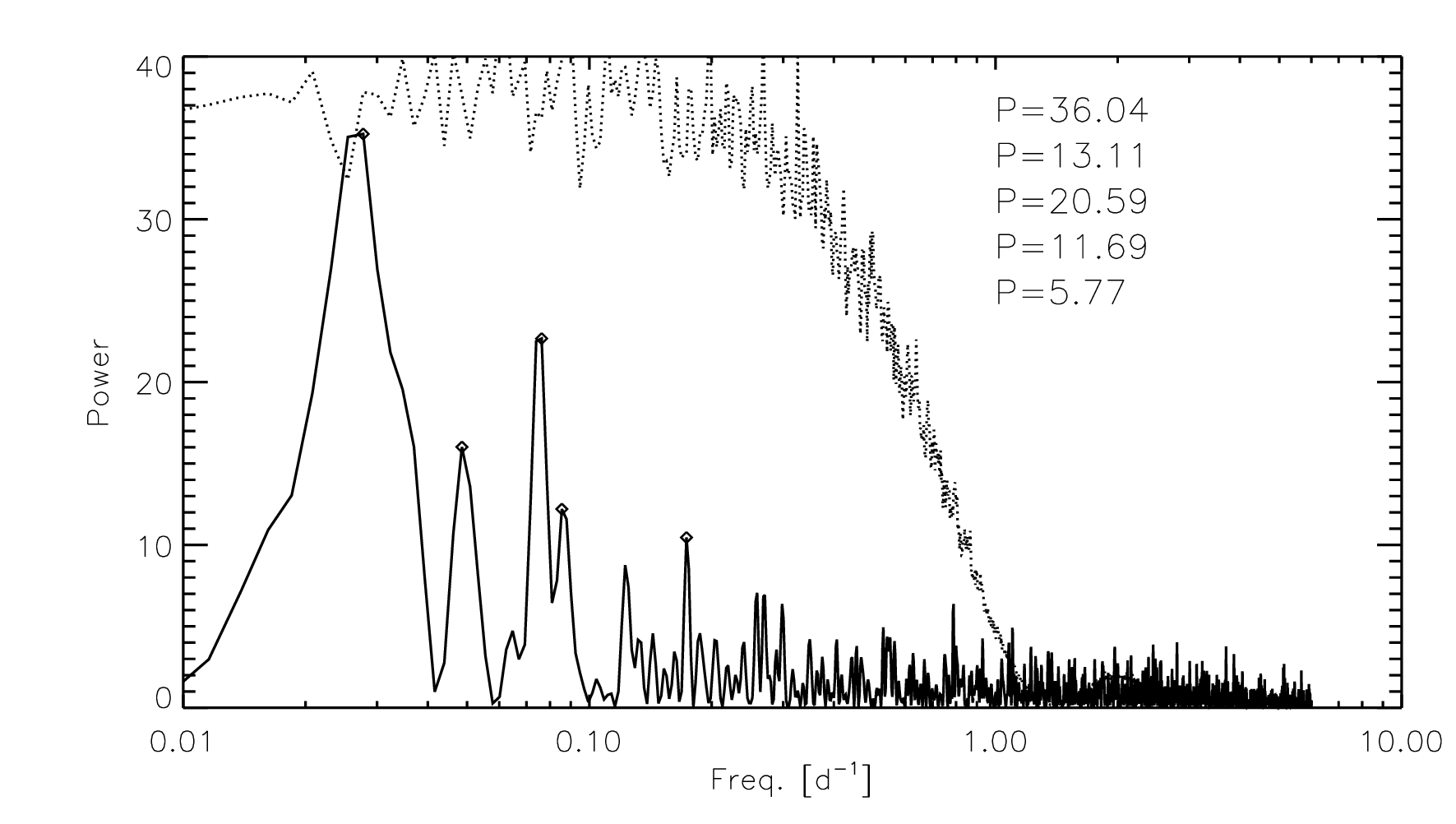}
   \includegraphics[width=8.cm,height=5cm]{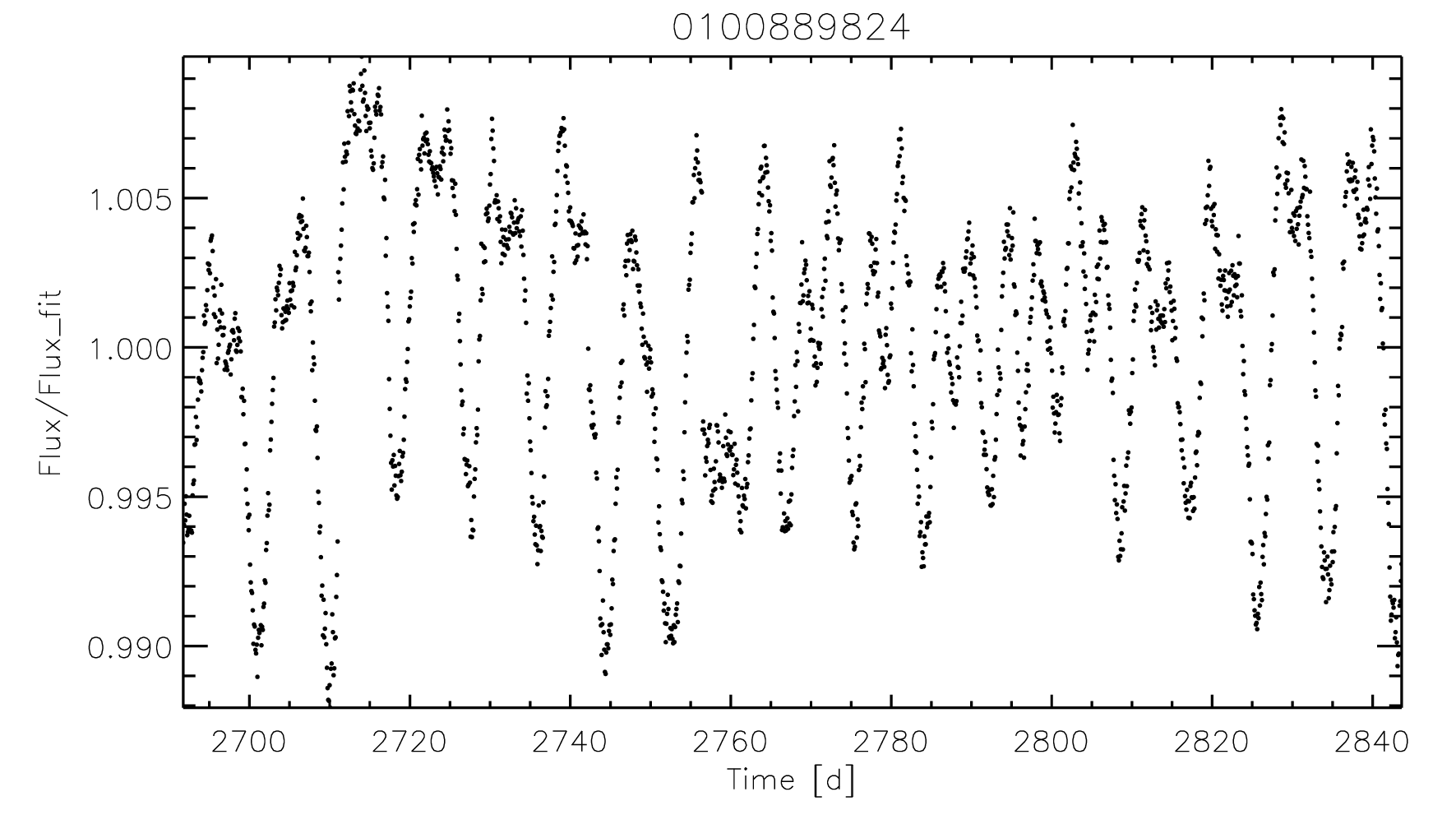}
   \includegraphics[width=8.cm,height=5cm]{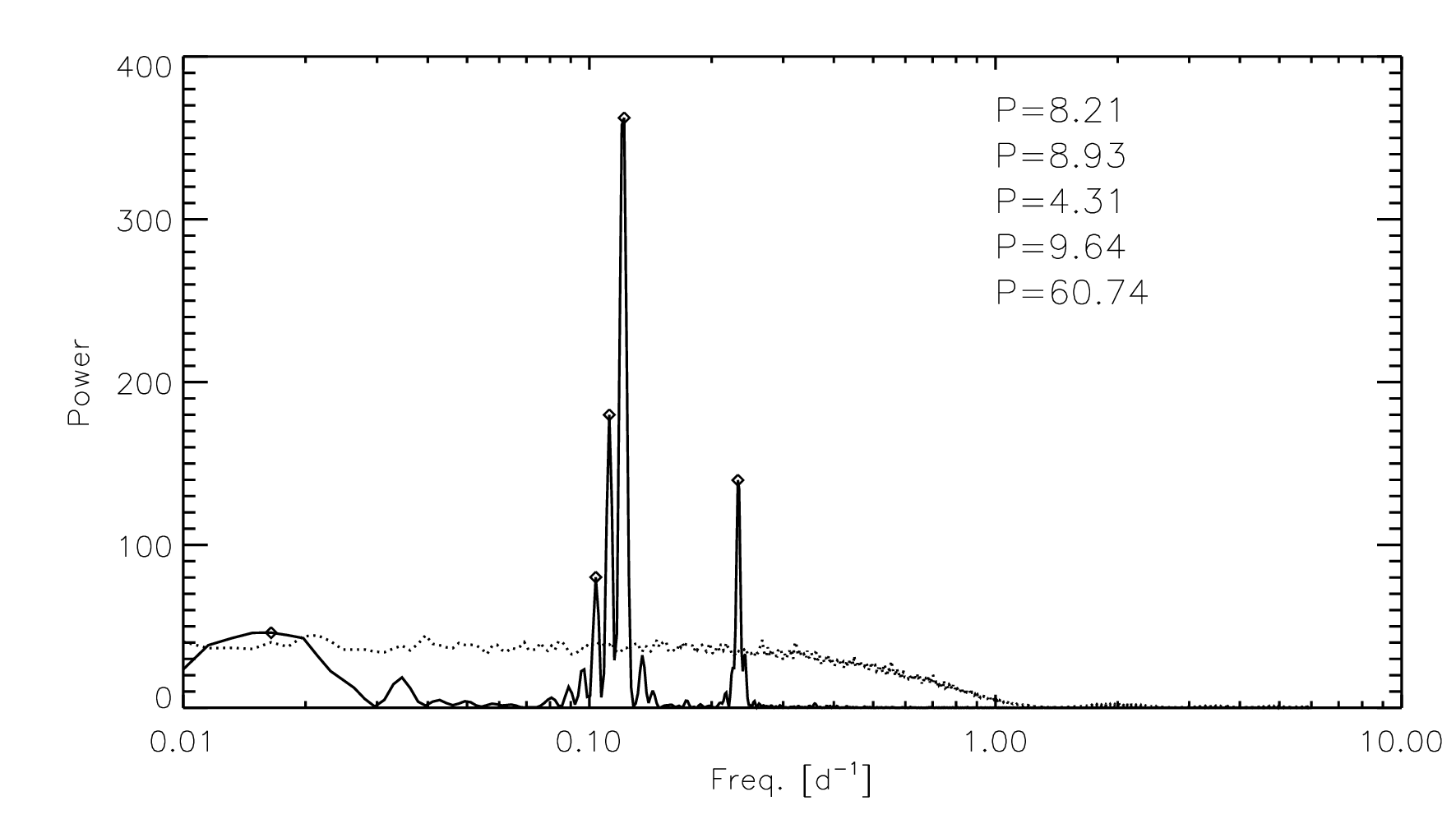}

   \caption{  Examples of the detrended light curves ($\it{left}$)
  for two CoRoT sources (observed towards the Galactic center);
periodograms ($\it{right}$) for the two LCs, the dotted curve
  superimposed on the LNPs is the threshold determined by simulations.
Periods found with the periodogram analysis for the star with CoRoT
  ID 0100993180 ($\it{top}$) are not significant since they have
P$_{orig-LC}$
  $<$ P$_{simul-LC}$, while periods found for the star 0100889824
($\it{bottom}$)
  are significant since they have P$_{orig-LC}$ $>$ P$_{simul-LC}$. Note the very different vertical axis.
              }
         \label{fig5}
   \end{figure*}

\section[]{Analysis of the CoRoT light curves}\label{analysis}

The analysis consists of  two different methods: the Lomb-Scargle periodogram analysis, the autocorrelation analysis and a consistency check supported by the visual 
inspection of each light curve. Our final result is the compilation of a catalogue of periods for field main-sequence stars, towards the Galactic center 
and anti-center. From a detailed analysis we determine which of these periods can be attributed to rotation. 

\subsection{Periodogram and folded light curves} \label{period}
We searched the light curves for the presence of significant periodicity using the
Lomb-Scargle Normalized Periodogram (LNP) approach \citep{lom76,sca82,hb86}.

The periodogram provides an approximation to the power spectral density of a time series. 
The approach used makes periodogram analysis exactly equivalent to least-squares fitting of sine curves to the data. 
A function $P(\omega)$ is computed from the data for a range of frequencies (which we have set from 0.01 to 10.0 d$^{-1}$, to approximately match the minimum and the 
maximum searchable periods in the observational window) and the frequency that maximizes this function is considered as
the most likely frequency in the data. 
Thus, we calculated the normalized power $P(\omega)$ as function of
angular frequency $\omega\,= 2\pi\,\nu$\, and identified the location of the highest peak in the
periodogram.\\ 
\indent In order to decide the significance of the peak we have followed  \citet{ehh95}, 
randomizing the temporal bins from the original
light curve. By
calculating the maximum power on a large number of randomized data sets, the conversion
from power to False Alarm Probability (FAP) can be determined. In detail,
we constructed 1000 light curves \emph{resampled} from the original ones randomizing the position of blocks of adjacent 
temporal bins (block length, 12 h) (e.g. \citealt{flaccomio05}). By shuffling the data we break any possible time correlation and periodicity of the light curve on time scales longer
than the block duration. 
We calculated the Scargle periodogram for all the randomized light curves and
we compared the maximum from the real 
periodogram to the distribution obtained from the 
randomized light curves, at the same frequency, in order to establish the probability that values as high as the observed 
one are due to random fluctuations. Given some threshold FAP$_*$\, we state that the detected candidate 
periodicity is statistically not significant if FAP $>$\, FAP$_*$. The calculation we performed on CoRoT light curves led often to small FAPs, 
indicating that LNPs of our light curves present peaks that in most cases cannot be explained by pure stochastic noise, or non-periodic variability on time scales shorter than 12 h. 
In fact, if light curves present variations on time scale smaller than the size of the temporal block we used in the simulations, these variations will be 
still present in the simulated curves and will be not recognized as significant. We
have chosen a bin size of 12 h as a reasonable compromise between the expected time scale of stochastic variations and the shortest expected periodic signal.

In particular out of the total analyzed sample of 6241 LCs, 4887 have significant periods (at FAP 1\%; since we simulated 1000 light curves, the 1\% FAP power is the power that was exceeded by the highest peak in 10 simulations). 
We show in Figure~\ref{fig5} some examples of the periodogram analysis.\\

\subsection{Autocorrelation analysis}\label{corr}

\begin{figure*}
   \centering
   \includegraphics[width=8.cm,height=5cm]{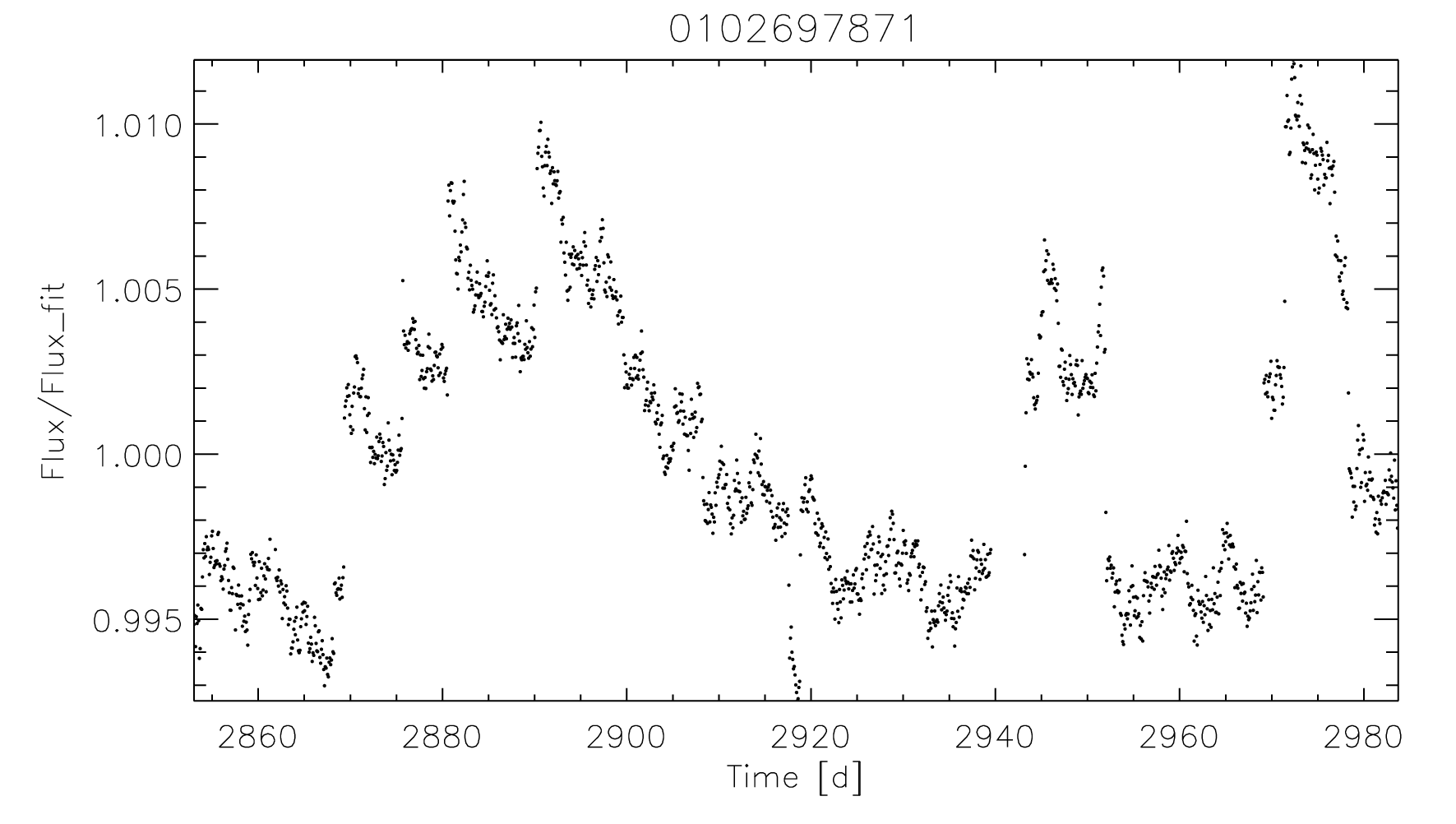}
   \includegraphics[width=8.cm,height=5cm]{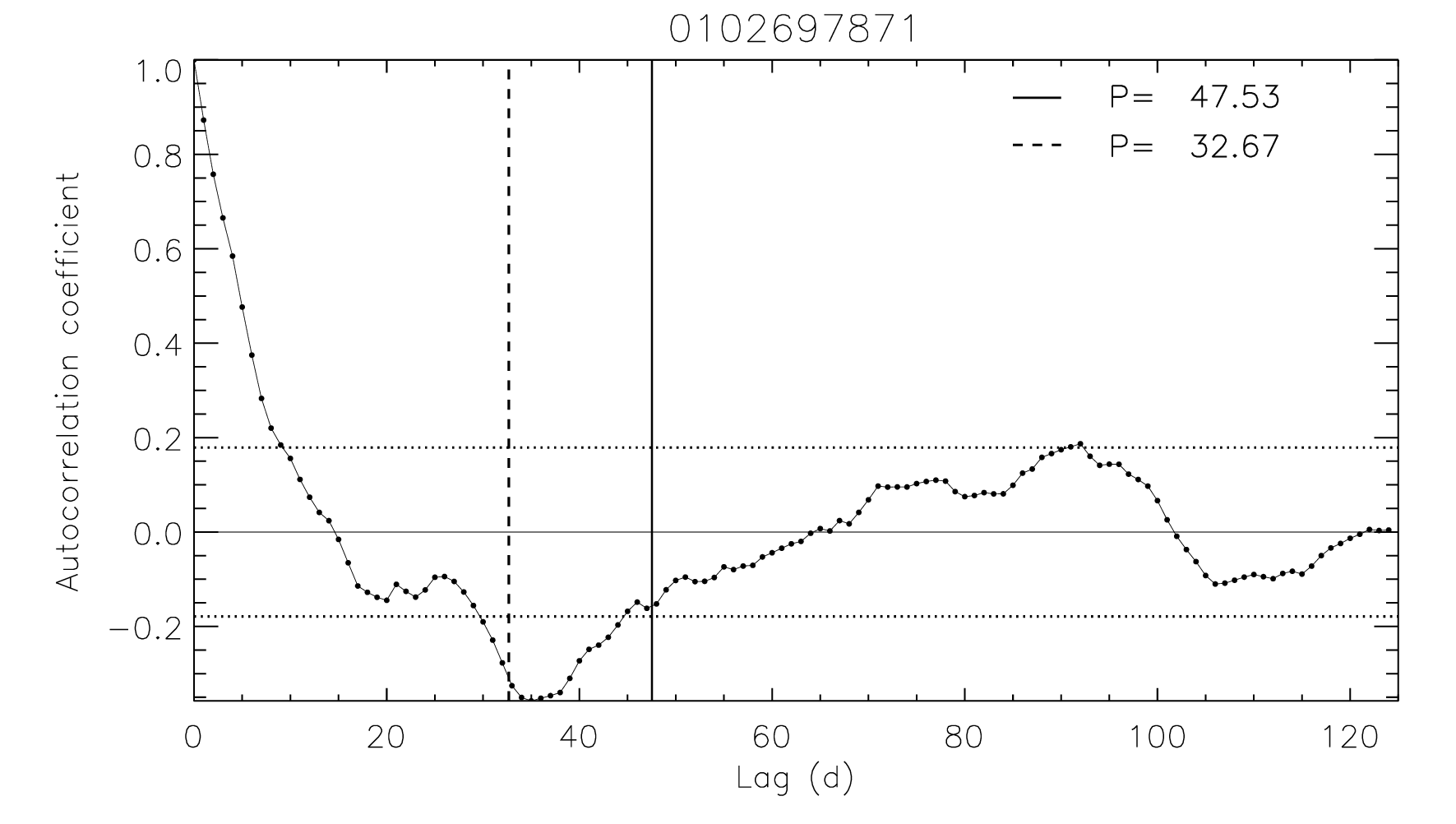}
   \includegraphics[width=8.cm,height=5cm]{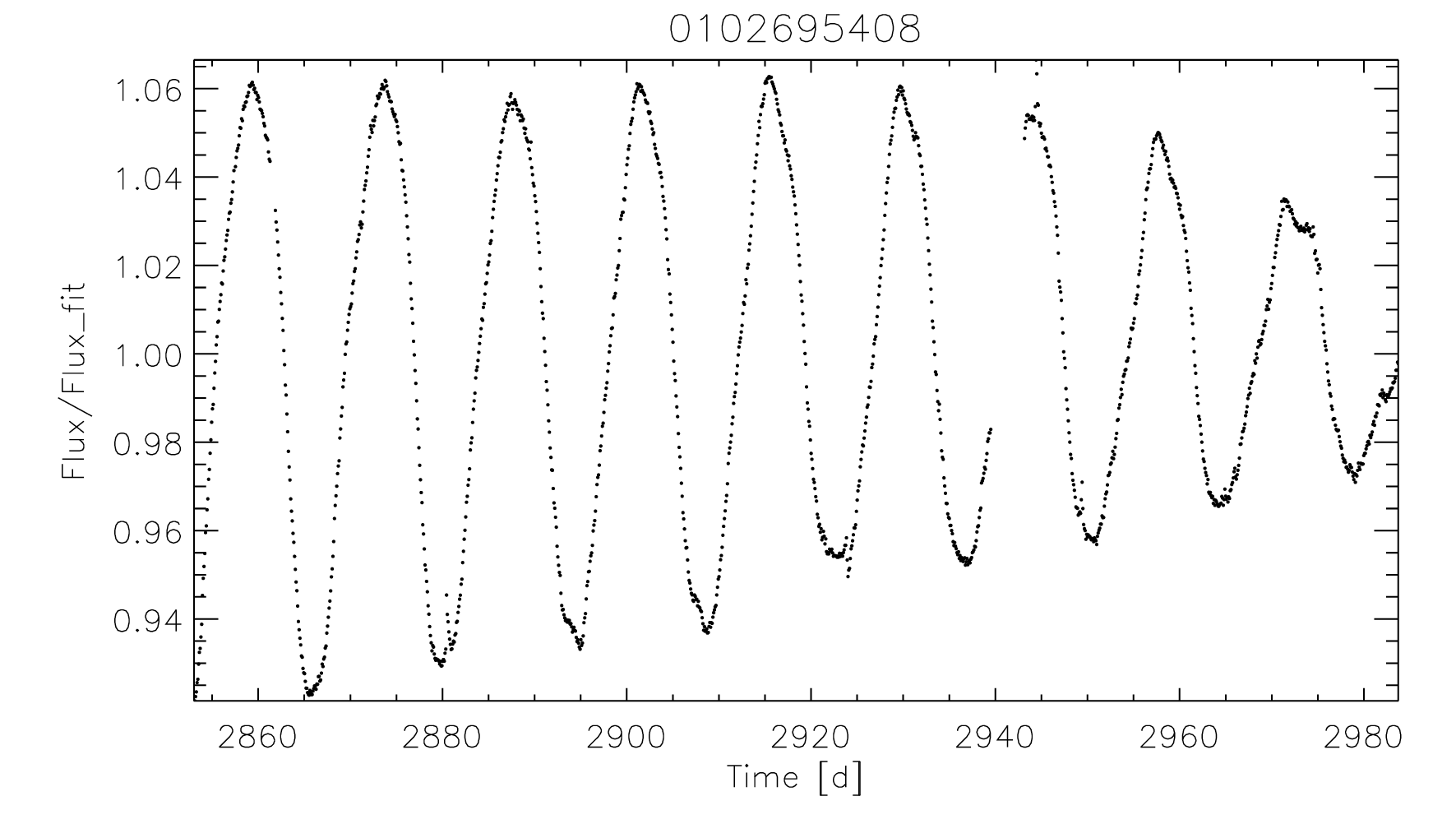}
   \includegraphics[width=8.cm,height=5cm]{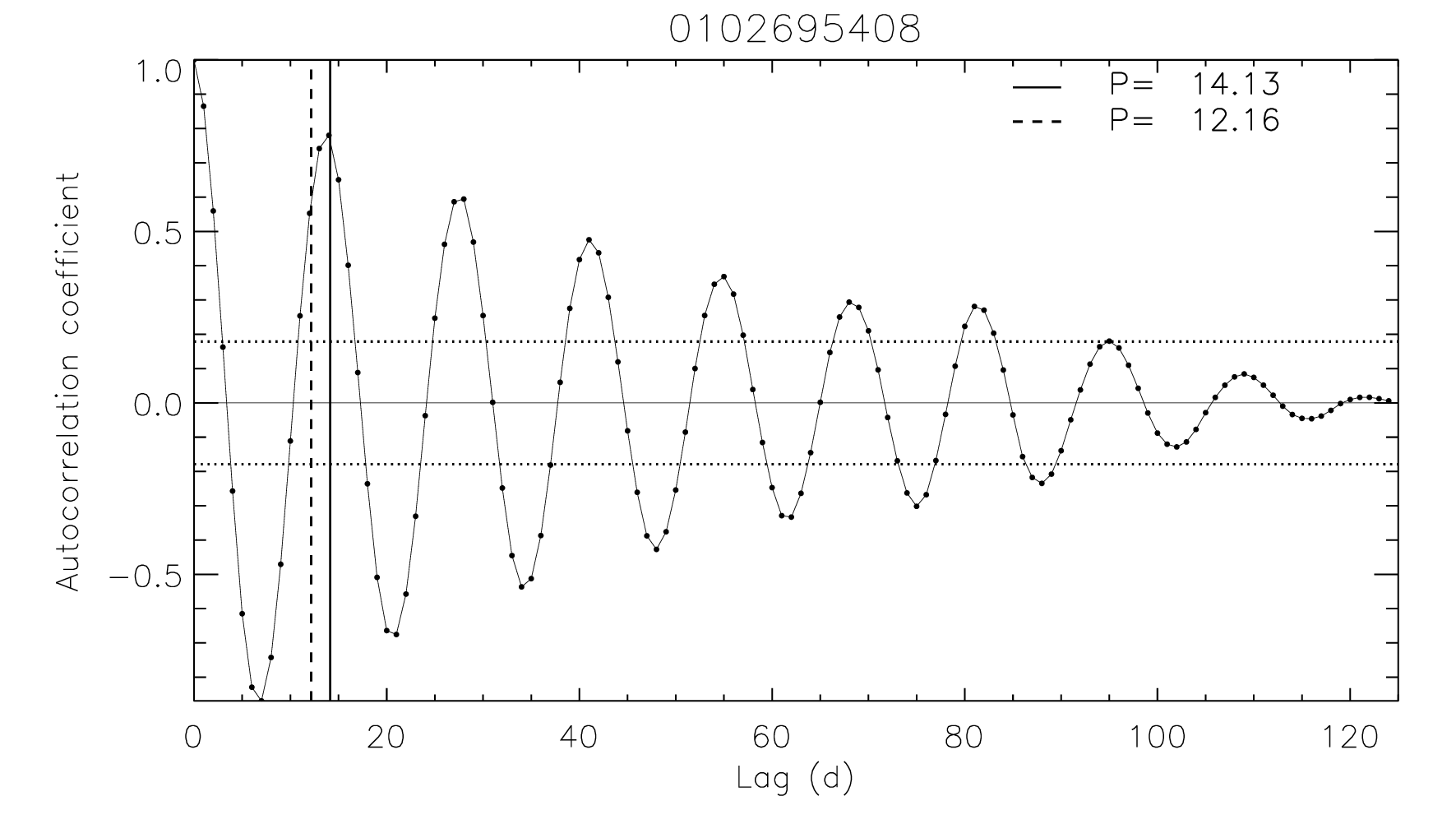}
      \caption{Two examples of light curves ($\it{left}\, panels$) and relative correlograms ($\it{right}$). In the
      autocorrelation plots we marked the 95\% confidence interval (dotted horizontal lines) and the first (solid vertical line) and second period
      (dashed vertical line) found by the periodogram analysis (the values of the first two periods yielded by the L-S periodogram are indicated in the top right of each plot). 
      }
         \label{fig6}
   \end{figure*}

Aliasing effects or residual effects due to the
particular choice of temporal blocks used in the light curve's simulation may cause spurious periods. To overcome these we used an
autocorrelation analysis \citep{box76}, mainly to confirm the periodicity of the
light curves, or to identify the correct period. Autocorrelation takes each point of the light curve measured at time $t$ and compares the value of that point
to another at time $t+L$. Points separated by $L$ will be very similar if the data contained some variability with period
$L$, thus the autocorrelation function will have peaks corresponding to periods of variability in the data. The
autocorrelation $r_{L}$ of a sample population X as a
function of the lag $L$ is:

$$ r_{L} =\frac{\sum_{k=0}^{N-L-1}(x_{k}-\bar{x})(x_{k+L}-\bar{x})}{\sum_{k=0}^{N-1}(x_{k}-\bar{x})^2}$$
where $\bar{x}$ is the mean of the sample population X and N is the sample size, the
quantity $r_{L}$ is called the autocorrelation coefficient at lag $L$. The correlogram for a time series 
is a plot of the autocorrelation coefficients $r_L$ as a function of L. A time series
is random if it consists of a series of independent observations with the same distribution. In this case we would expect the
$r_L$ to be statistically not significant for all values of $L$. 
We have chosen to adopt a 95\%  confidence level to select significant autocorrelation coefficient. 
Following this criterion a total of 3578 LCs present a significant autocorrelation.\\
\indent In Figure~\ref{fig6} we show two examples of autocorrelation
plots for two stars of our sample. In this figure we can notice the particular behaviour of the top LC, the one with CoRoT ID 0102697871. The L-S periodogram found three significant periods ($P_1$\,= 47.5, $P_2$\,= 32.7,
$P_3$\,= 24.9 d) at variance with the correlogram, in which no significant period is found, moreover the phase foldings performed with the three periods indicated by the L-S method show a very large scatter. We analyzed in detail this star in order to understand its behaviour and discriminate between astrophysical and instrumental signals. In particular, we compared the evolution of the
flux in the three different channels (red, green and blue) which were available for this star, and which show that instrumental signatures significantly perturb the LC. There is a data gap of about 4 days between 2939 and 2943. A
proton impact led to a reset of the Data Processing Unit (DPU) 1, which is responsible for data collection on the E1 CCD, on January 18, 2008 during SAA crossing. All light curves in LRa01 originating from CCD1 contain this data gap.
The stepwise periodic changing between 2870 and 2900 is not real but it is due to a peculiar combination of hot pixel of high and low amplitude affecting different channels. In particular, in $T=$\,2858, 2867 two small amplitude hot
pixels appeared in the blue channel, and two bigger events at $T=$\,2870 and 2890, then two big ones affected the red channel at  $T=$\,2876 and 2882, the rest of the up and down behaviour is due to strong hot pixel events in only
the blue channel. The red and the green channel display opposite trends (as usual) with the red channel affected by hot pixels in the first 30 days of the LC, thus also the large variation shown by the white curve seems to be
instrumental. There are several light curves displaying a similar behaviour as the illustrated one, and we performed an analogous analysis using colour channels, when available. In other cases, with no colour information, we had to
trust the cross-correlation of the results obtained with L-S periodogram, phase folding, autocorrelation technique and visual inspection, but always following a conservative approach which minimizes the number of spurious periods in the final catalogue. 

\subsection{Consistency check and visual inspection}\label{visual}

After having  derived periods both from the periodogram  and the autocorrelation analysis, we have checked the consistency of the results obtained with the two methods.
In none of the  cases in which the Lomb-Scargle method does not find any significant period,  the autocorrelation method finds  one, that implies that the latter is more conservative 
in identifying periodicity.
A total of 1457 stars have consistent periods according to both methods.
In 1152  cases  we have periods significant for the Lomb-Scargle analysis  but not significant for autocorrelation. Since we want to derive a sample with confirmed periods, we have 
discarded these light curves. 
In 2121  cases both methods give discrepant significant periods. We have  examined by eye the phase folding with both periods, choosing as  
final period the one that reproduce a smoother behaviour. In many of these cases the rejected period is an alias, and the folding helps in discriminating the 
original period (Figure~\ref{fig7}, left). This approach has made possible to determine periods for 521 LCs , in the remaining cases the folding gave unsatisfying 
results (very scattered plots) and therefore we discard them from the periodic LCs. We want to stress that there may be a complex mixing of several
phenomena (i.e. chaotic activity, fast evolution of spots, multiple periods which interfere with one another, residual instrumental effects) which makes the identification of a clear periodicity extremely difficult, on the basis of
the photometric behaviour only. For these stars it would be useful to analyze both LCs collected at different
times and spectroscopic observations. In a large fraction of these discarded periods, we noticed the presence of 
several small jumps, that are not identified by our procedure described in Sec. \ref{red}, that strongly affect the analysis producing spurious 
periods. 
An examples of such cases (spurious period) is shown in Figure~\ref{fig7} (right).\\
\indent{In total the described procedure has allowed us to determine periods for 1978 stars (1457+521).

\begin{figure*}
   \centering
   \includegraphics[width=8.cm]{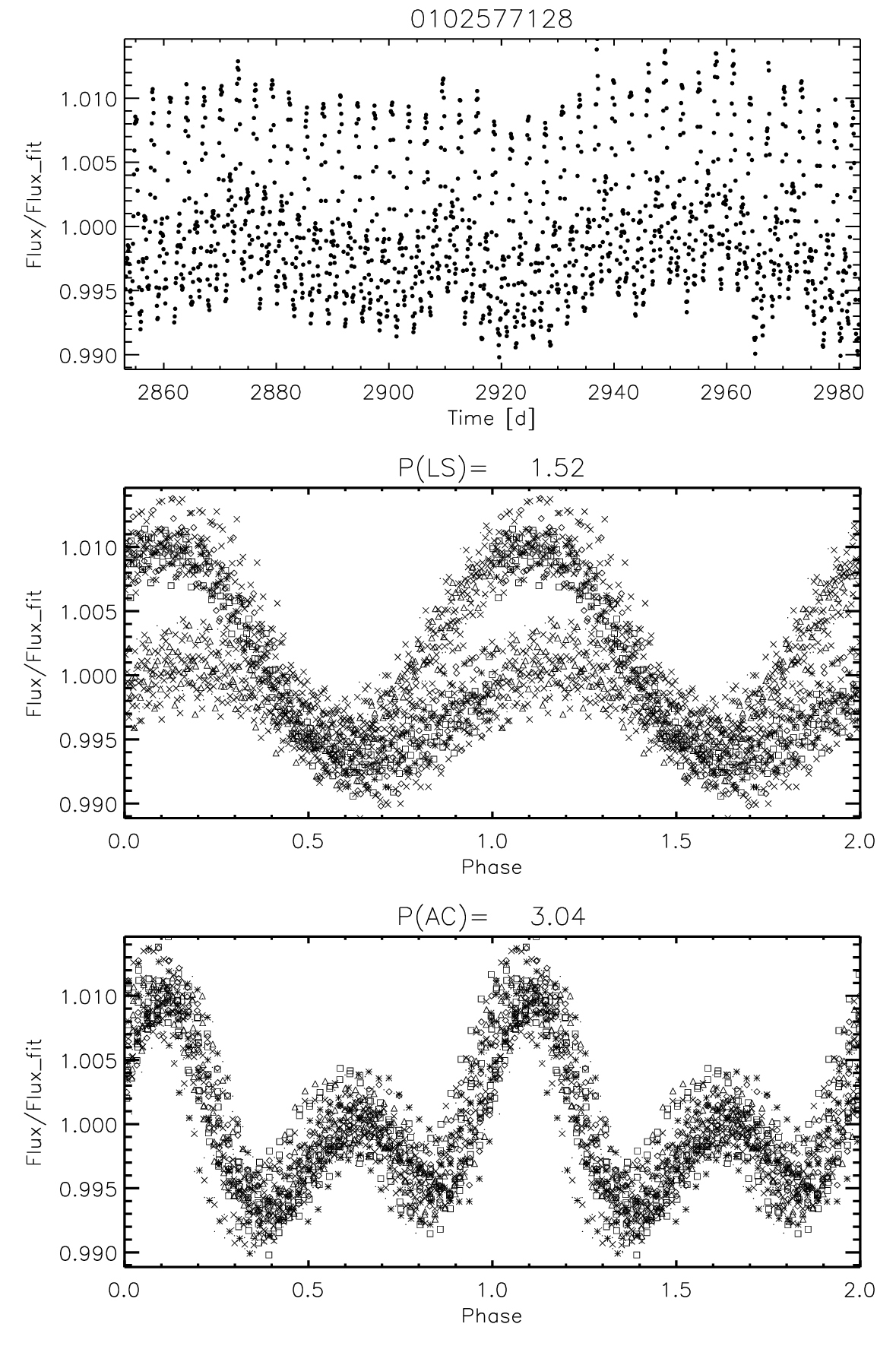}
   \includegraphics[width=8.cm]{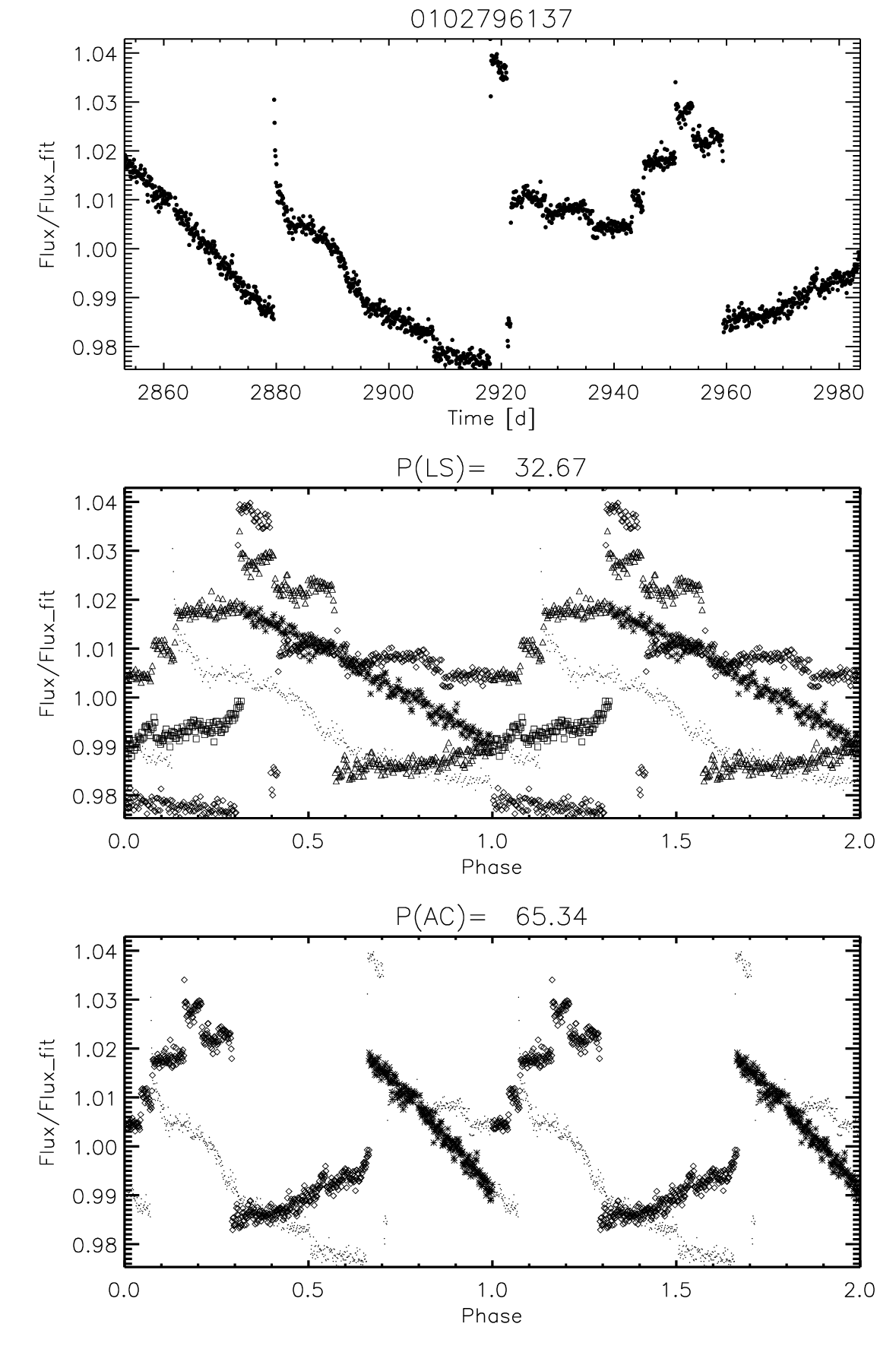}
\caption{In the $\it{left}\, panels$ is shown an example of a LC ($\it{top}$)
  for which the most significant period derived with the
  Lomb-Scargle method P(LS) is an alias of the period indicated by the
  autocorrelation analysis P(AC), as can be seen in the relative phase
  foldings ($\it{middle}$\, and $\it{bottom}$\, panels, respectively).
  In the $\it{right}\, panels$ is shown an example of LC ($\it{top}$) affected by several
  jumps (smaller than the 10 $\sigma$ threshold). In this case the two methods find discrepant periods. 
  On the basis of the analysis of the relative phase foldings ($\it{middle}$ and $\it{bottom}$ panels) we decided to discard these periods.
         }
               
         \label{fig7}
   \end{figure*} 

\begin{figure*}
   \centering
   \includegraphics[width=8.6cm]{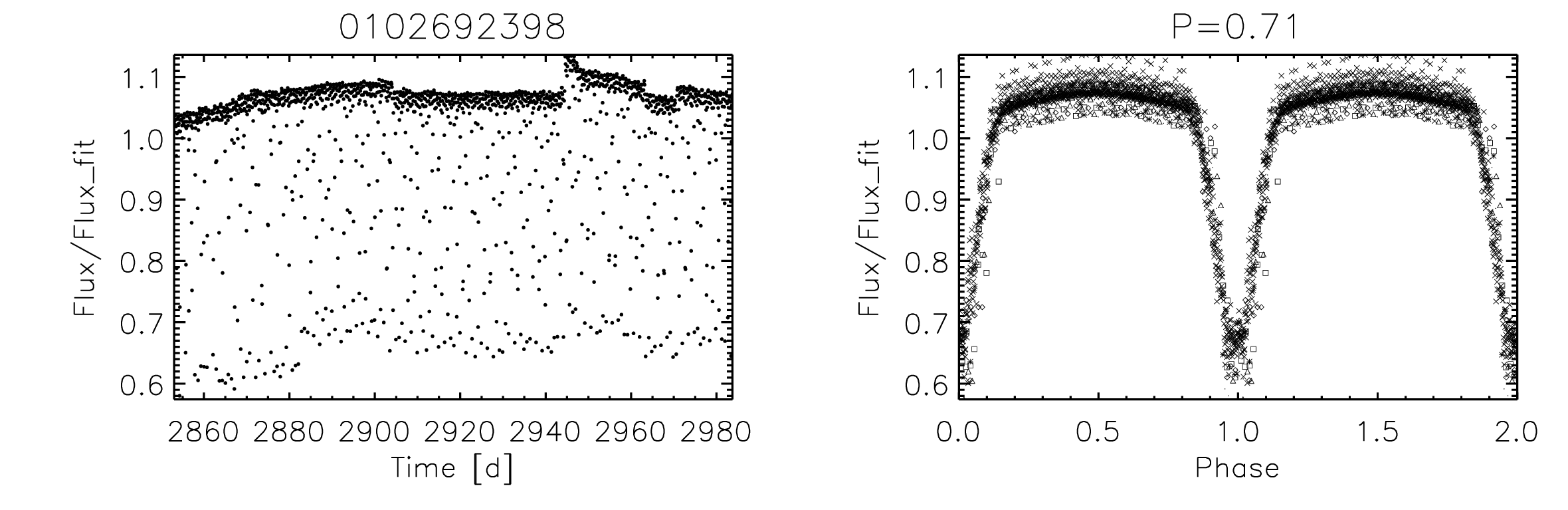}
   \hspace{0.1cm}
   \includegraphics[width=8.6cm]{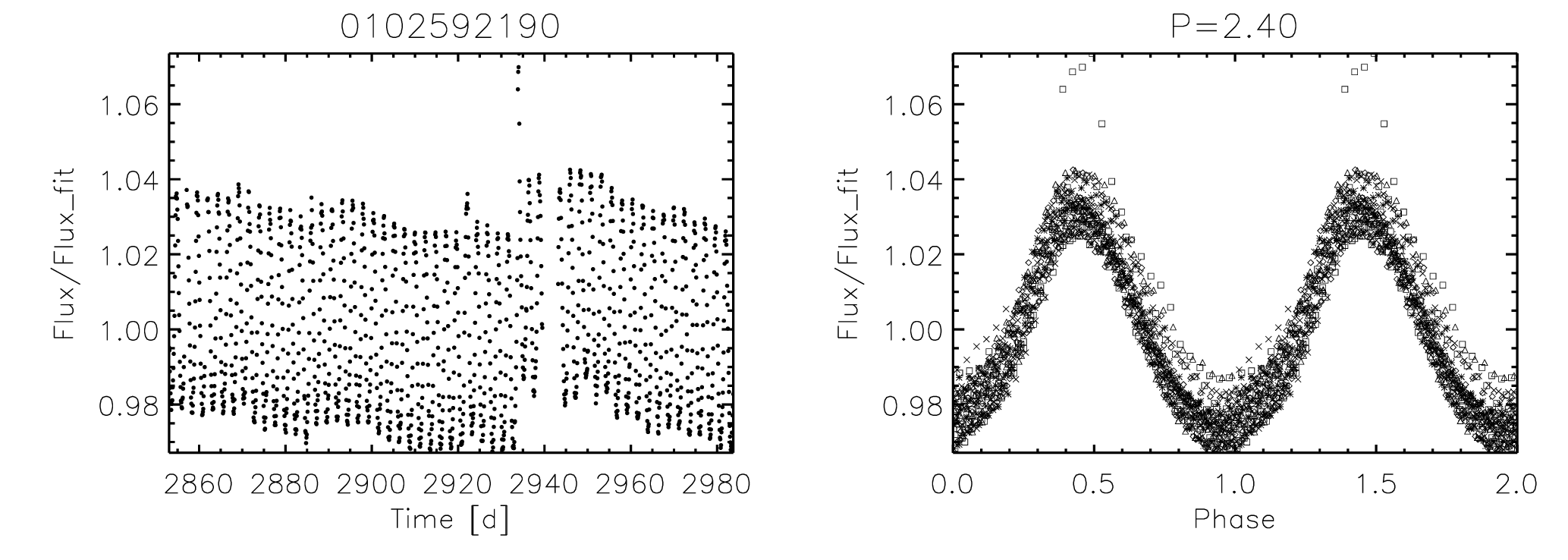}
   
      \caption{Two examples of light curves and folded LCs  (first Lomb-Scargle
     period) for an eclipse star
     (ID: 0102588918, $\it{left}\, panels$) and for a pulsator (ID:
0102592190,
     $\it{right\, panels}$).
            }
         \label{fig8}
   \end{figure*} 
   
\section{Results: Identification and analysis of rotational periods}\label{signif}

In order to study the rotational properties of our sample we need to identify the eclipsing systems and pulsators. 
Eclipsing stars are easily identified by eye from the narrow eclipsed region in the folded curve (see an example in Figure~\ref{fig8}, left panels). 
We identify a total of 82 eclipsing systems. However in a fraction of them it is possible to derive both the orbital and rotational periods. 
In these cases we report both periods in Table 2.

A further step is the identification of pulsators (see for example Figure~\ref{fig8}, right panels). \\ 
\indent The distinction between pulsations and rotational modulation is difficult to make solely based on the light curve.
To distinguish  between variations due to pulsation and rotation caused by spots we adopted a method which
exploits both the stability of the detected periods for long time and the small flux dispersion in the
folded curve \citep{sjl+04,sjl+05a,sjl+05b}.\\
\indent Various observable parameters have been proposed to distinguish between different types of pulsators. \citet{wb08} define pulsators as stars displaying regular pulsation brightness variations with periods less than 2 days. The
majority of these pulsators displays regular radial sinusoidal variations. Longer period pulsators are defined as stars displaying regular sinusoidal or slightly sawtooth variations with various periodicities greater than an
arbitrary chosen lower limit of 2 days. These stars are likely giant stars of late spectral types. Apart from the definition used and the particular form of the LCs, one distinctive characteristic of pulsators is the longer term
stability of the pulsation \citep{sjl+04,sjl+05a,sjl+05b,dsl+09}.\\
\indent We adopted a criterion based on the hypothesis that pulsator LCs are much stabler than rotation LCs. Then we divided the phase interval in 40 bins and we calculated the median flux value for each of these bins. Through linear
interpolation of these median values we derived the median flux as a function of phase (see Figure~\ref{fig9}) and we calculated for each phase value the 
ratio between the difference of the instantaneous flux and the median phase folded lightcurve at the corrisponding phase and the peak-to-peak 
amplitude variation of the LC, Flux$_i$-Flux$_{interp_i}$/(Flux$_{max}$-Flux$_{min})$.
We identify a star as a pulsator if  more than 80\% of the Flux$_i$-Flux$_{interp_i}$/(Flux$_{max}$-Flux$_{min})$
ratios are lower than 0.15. These 
thresholds have been chosen after a detailed inspection of all the LCs, identified with pulsators by eye (LCs very regular with very small dispersion), and of a comparison sample of clearly non-pulsating LCs. 
The described method retrieves about 90\% of the pulsators identified by eye. To be conservative,  we add the remaining 10\% LC to the pulsator sample.  We have decided to not increase the chosen 
threshold to include them, since otherwise we would include a significant number of LCs with evident characteristics of rotators (more irregular and higher dispersion LCs).
A total of 169 pulsators has been found, 111 of which have periods smaller than 2d. \\
\indent \citet{dsl+09} performed an extensive automated classification of variable stars for the first four CoRoT exoplanet fields, stating that we do not expect to find many classical radial pulsators while more candidate multiperiodic, non-radial,
pulsators are found, due to the selection of the CoRoT exoplanet observation fields biased towards cool main-sequence stars. Several stars identified as pulsators with our method are probably giant stars, contaminating our sample.
Amongst selected pulsators we recognized several double-mode RR-Lyrae, few low-amplitude Cepheid pulsators and a few RR-Lyrae displaying a clear Blazhko effect (the LC varies in height and shape from cycle to cycle due to still poorly
understood processes which affect the regular pattern).\\ 
\indent We compared our classification of pulsators with the one made by \citet{dsl+09} for the common objects and we found that, using our method, we always succeed in making the same
classification as the \citet{dsl+09} one.\\ 
\indent Amongst the LRa01 identified pulsators we found 16 object which were also observed during the
IRa01 (the initial run toward the galactic anti-center, with a data lenght of 55 days, which partially overlaps with the LRa01 field). As a further verification, we analyzed the LCs for these objects, the periods found for the two
sets of data (IRa01 and LRa01) are the same, and we classify these 16 IRa01 objects also as pulsators, following our method. There is a gap of $\approx$\,206 days between the two observing runs, thus we phased the two sets and we found
that the pulsations are stable during this longer range of time. Such long time stability of periods is highly improbable for spot-like modulations. As \citet{dsl+09} admitted, rotation can produce LCs that are difficult to
distingush, for example, from the typical skew-symmetric Cepheid LCs, nevertheless we need spectral information and LCs with a longer total time span to verify whether the detected periods remain stable over time.  \\ 

\begin{figure}
   \centering
   \includegraphics[width=8.5cm]{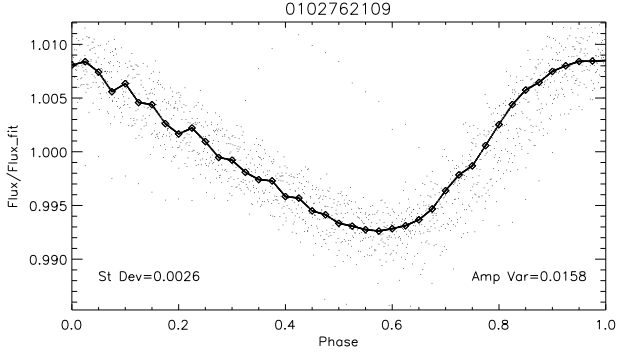}

      \caption{Interpolation curve of the
median folded flux values for a candidate pulsator.}
         \label{fig9}
   \end{figure} 

We remain with 1727 stars whose modulation is interpreted as due to rotation. 
All the derived periods and their classification are reported in Table ~\ref{tab01}.
The table shows a portion of the catalogue: the first and second columns report the CoRoT and 2MASS ID of the stars; the
third and fourth their right ascension and declination; the 5th column reports the derived periods; the 6th and 7th are flags indicating the method(s) adopted to derive periods and
the nature of the stars (pulsator, eclipse star, etc), respectively and the
8th is a link to the plots of the detrended light curve and of the folding.  The 9th column reports the rotational periods of eclipsing binaries, when derived.

The distributions of rotational periods for F, G and K-type stars (both Galactic anti-center and 
center) are  shown in Figure~\ref{fig10}, the distribution of periods of F stars in the center direction is missing because of their paucity, only 4 stars and none with significant period.

The distributions shown in Figure~\ref{fig10} are bimodal, with two peaks probably associated to 
two distinct populations, one (presumably young stars and binaries) peaking at very short rotational periods
(P$_{\rm rot}$~$<$~5 - 10 days), and the other (older) peaking at longer rotational periods,  with a  
gap between about 15 and 35 days.

The presence of a large number of stars with short rotational
periods is fully compatible with the presence in the solar neghborhood  of a sample of young stars, both 
towards the center and the anti-center of the Galaxy. 
Stars in the short period peak in Figure~\ref{fig10} have periods shorter than typical for old single 
stars. They are likely dominated by young stars that have not had the time to spin down with 
perhaps a 
fraction of older tidally locked binary stars. In fact, by visual inspection of the light curves we have 
been able to detect some
binary systems of different type (eclipsing, contact and detached binaries).

\begin{table*}
\centering
\caption{Catalogue of periods for CoRoT main-sequence stars towards the Galactic center and anti-center (the complete table is available online).}
\label{tab01}
\begin{tabular}{cccrrcccc} \hline
CoRoT ID   &  2MASS ID           &RA$^1$         &  DEC$^1$       & PERIOD (d) & METHOD$^a$  & LC TYPE$^b$   & PLOTS & NOTE$^c$\\\hline
0102571157 & $06403718-0019246$  & 100.15499     &   -0.32350     &   7.261  &    1-2   &  R &  {\tiny 0102571157.pdf}& -- \\
0102577348 & $06404793+0009568$  & 100.19977     &    0.16580     &  18.027  &    2     &  R &  {\tiny 0102577348.pdf}& --\\
0102680285 & $06431422-0018127$  & 100.80924     &   -0.30352     &   6.150  &    1-2   &  P &  {\tiny 0102680285.pdf}& -- \\
0102640143 & $06422344+0056313$  & 100.59766     &    0.94205     &   1.815  &    1-2   &  P &  {\tiny 0102640143.pdf}& -- \\
0102629072 & $06420795+0043453$  & 100.53311     &    0.72926     &   3.901  &    1-2   &  E &  {\tiny 0102629072.pdf}& --  \\
...&...&...&...&...&...&...&...&...\\
...&...&...&...&...&...&...&...&...\\\hline
\end{tabular}
\begin{flushleft}
$^1$: J2000.0\\
$^a$: 1: Periodogram analysis; 2: Autocorrelation\\
$^b$: P: Pulsation; R: Rotation; E: Eclipsing binary\\
$^c$: Rotational periods of eclipsing binaries
\end{flushleft}
\end{table*}

\begin{figure}
   \centering
   \includegraphics[width=8.5cm]{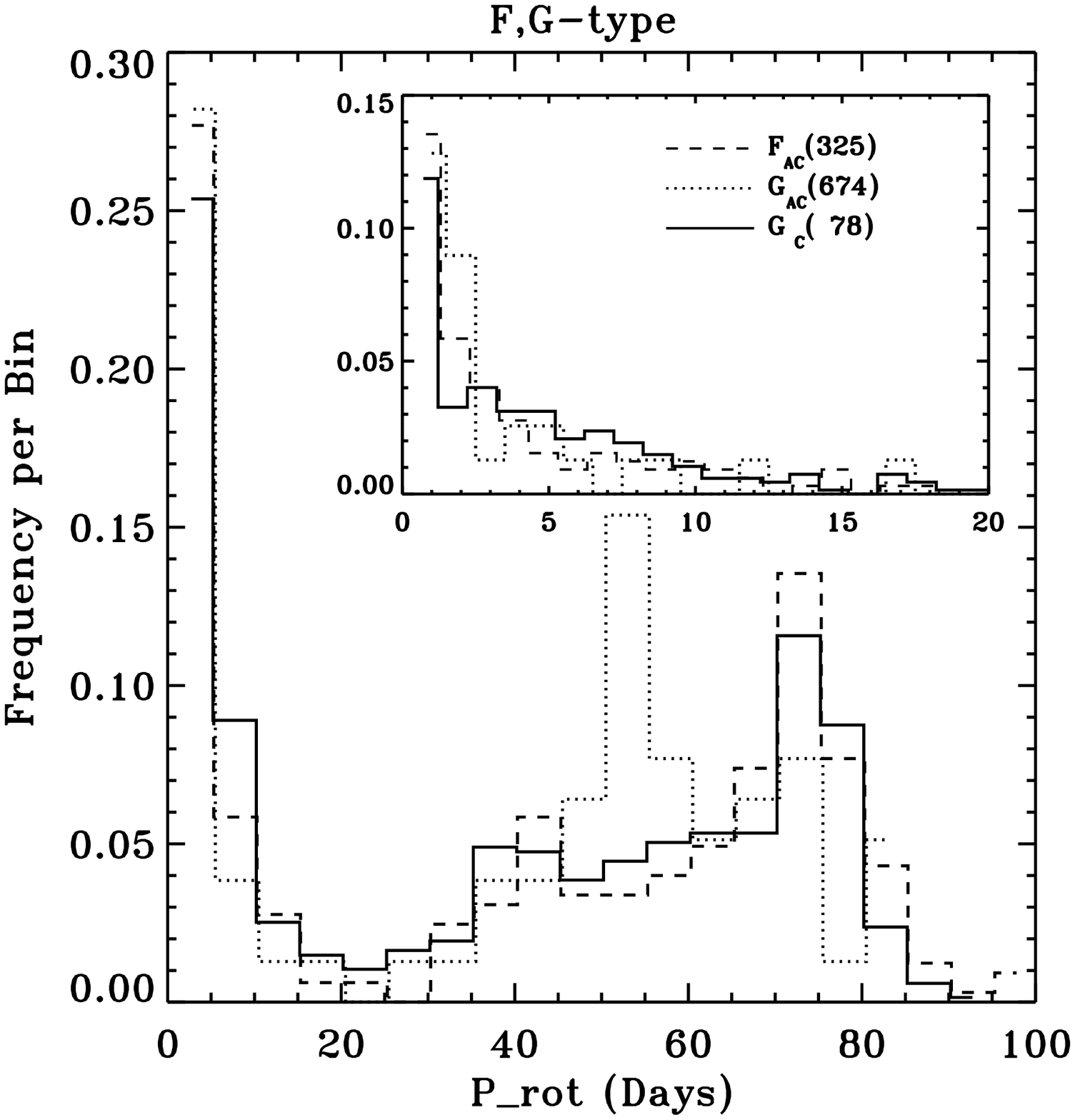} 
   \includegraphics[width=8.5cm]{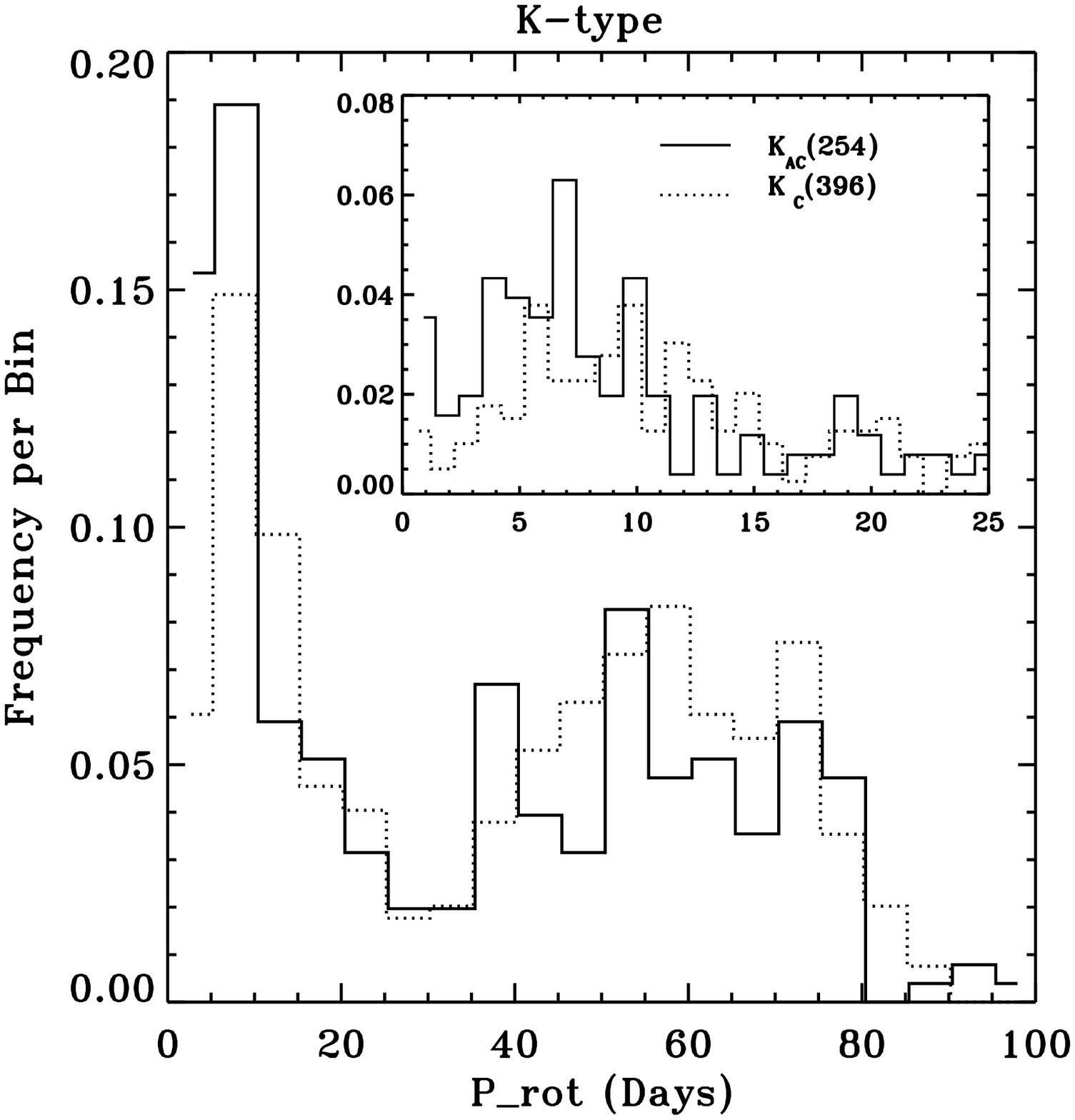} 
      \caption{Normalized distributions of rotational periods for the 
      sample analyzed of F and G-type stars ($\it top\, panel$) and K-type ($\it bottom\, panel$) stars towards the Galactic anti-center and center. 
      The distribution of periods of F stars in the center direction is missing because of their paucity (4 stars). The numbers reported indicate 
      the number of light curves for each type.
              }
         \label{fig10}
   \end{figure}  

\begin{figure}
   \centering
   \includegraphics[width=8.5cm]{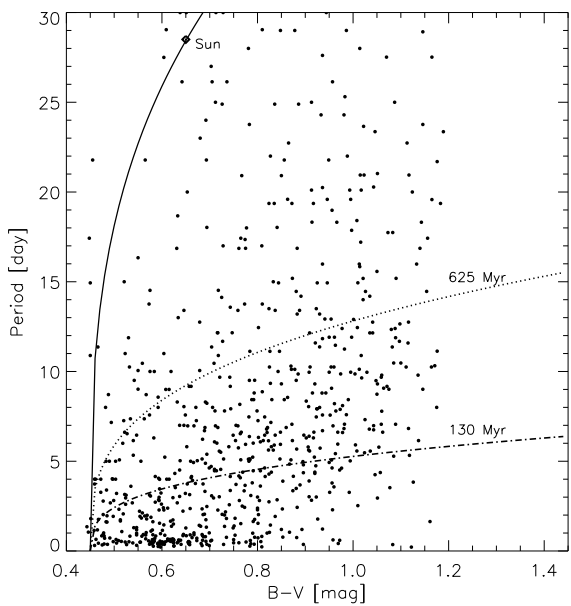} 
      \caption{Rotation period vs $B-V$ derived from CoRoT light curves for solar-type stars compared to 
      the relations derived in \citet{mh08} for 130 Myr (Pleiades) and for 625 Myr (Hyades). 
      }
         \label{fig11}
   \end{figure}

\section{Discussion}\label{discuss}
As described in the introduction, rotational period may be used as a proxy for age for single dwarfs of a 
given mass.  In particular this is possible for stars older than $10^9$ years when solar type stars 
have converged to a single value rotation-age relation. 
This has been confirmed by a 
number of studies of rotational rate in intermediate-age open clusters (e.g. the 
150 Myr old M35, \citealt{mms09}, the 250 Myr old M34, \citealt{iah+06}, the 550 Myr old M37, \citealt{mdp+08}, the 600 Myr old Coma cluster, 
\citealt{ccd09} and the well studied 625 Myr old Hyades cluster, \citealt{rtl+87}) which have shown 
that the rotation of F and G stars at the age of 150 Myr is already 
single-valued and only dependent on color (that is, mass) and age, \citep{mms09}, while for K-type 
stars this happens at the age of the Hyades, at which rapid rotators are still 
present among M-type stars. We refer the reader to the recent work of \citet{ibb+11} (and references therein), for a detailed discussion on rotation and angular momentum evolution of M-type field main-sequence stars from the MEarth
transit survey. They measured a wide range of rotation periods for 41 M dwarfs (from 0.28 to 154 days), finding that kinematically young (thin disk) objects rotate faster than the kinematically old (thick disk) objects.  \\
\indent \citet{bar07} used literature data to derive the gyrochronology relation.  According to him, the 
rotation periods evolve with age as: 
$$ P(B-V,t) = f(B-V) g(t),$$ showing a dependence on both age and color.

Since our sample includes very few dM stars, we may safely consider that stars slower than the 
Hyades rotation-mass relation, are older than 600 Myr. Faster stars are certainly younger. 
Therefore we used the Hyades rotation-mass  relation \citep{bar07,mh08}, as a dividing line 
between stars younger and older than 600 Myr, as shown in Figure~\ref{fig11}. Gyrochronology relations for aged main-sequence stars 
rely only on the Sun and few other stars, thus, even if they are older, we decided to exclude rotational periods
greater than 30 days from the plot.\\ 
\indent As can be seen in the $P_{rot}$ vs $B-V$ plot, more than 500 stars rotate faster than the mean rotation rate expected for the Hyades, and we consider these stars as dominated by a young population ($\le 600$
Myr). Our sample suffers of several biases  preventing us to perform a comparison with the observations 
and the predictions from Galactic models. In fact the original sample was selected to maximize the 
probability to detect planetary transits, on top of that  we  discarded from our analysis  a fraction of 
targets affected by instrumental artifacts, and finally we have not been able to identify rotational 
periods 
in a significant fraction of stars. There are several reasons for that, the most obvious is that we need 
a minimum  level of activity to detect a minimum rotational modulation. This will bias the final 
sample against the more quiet stars, likely the oldest ones. On the other extreme, chaotic  activity, 
characteristics of very young stars may prevent the identification of a clear periodicity. 

Stellar rotation is one of the best  
proxy for young main sequence stars, since their photometric and spectroscopic properties do not differ significantly from those 
of the oldest field stars. However this population may be contaminated by close 
binary systems, which have maintained higher rotation rates due to tidal synchronization. 
We can estimate the fraction of 
binary stars in our samples from the work of 
\citet{dm91} and \citet{mdh+92}, on the statistics of binaries in the solar neighbourhood. \citet{dm91} 
found that the fraction of binaries among G-type stars in the solar 
neighbourhood is about 60\%, 6\% of which is constituted of short period binaries, those of interest here 
(P~$<$\,~20 days). 
\citet{mdh+92} found that the analogous fraction in K-type stars is about 45\%, with a contribution of 
about 3\% from short period binaries. These estimates imply that binaries may account only for a small fraction of our fast rotator sample, then we can 
safely identify the bulk of  our fast rotators  as composed mainly of single stars.

Previous studies based on the stellar content of shallow and intermediate X-ray surveys have 
revealed a discrepancy between
model star counts based on constant rates of star formation and observed star counts for active 
stars. In particular, an excess of X-ray bright FGK stars has been observed \citep{fsr+88,sfm95,lms+07}, for example,
by  the \textit{Einstein} Extended Medium Sensitivity
Survey (EMSS, \citealt{gms+90}),   the \textit{XMM-Newton} Bright Serendipitous Survey (XBSS, 
\citealt{dmc+04})  and the 
\textit{ROSAT} North Ecliptic Pole survey \citep{hgm+01}.
These late-type stars appear to be young, as shown by their lithium abundance \citep{fbm+93}, 
although a fraction may be active binary
systems that have a yellow dwarf as a primary \citep{lms+07}. Data  are consistent with the 
presence of a recent event of star formation in the last $10^8$ years, while 
in the range $10^8$-$10^{10}$ years, the assumption of a model with a constant rate
of star formation agrees with the observations \citep{amm08}.

Analogously several young (8 - 50 Myr) moving groups have been identified both in the northern 
and southern sky  (\citealt{zs04}, \citealt{lmc+06} and reference therein), suggesting that the history of the star 
formation close to the Sun may be very complex with a few recent bursts.
 
In  light of the mentioned suggestions, reported in the literature, of a recent event of star formation 
in the solar neighbourhood, it is plausible that the fast rotators identified  with our analysis  
trace
this young population. Spectroscopic observations of these stars would clarify their nature and 
provide a further test of the presence in the solar neighbourhood of the young
population indicated by the activity surveys.  

\section{Summary}\label{concl} 

We have analyzed the light curves observed by CoRoT in  two fields at different direction in the Galaxy, in order to derive stellar 
rotational periods. Our original sample of 8341 light curves has been reduced to 6241 due to 
instrumental artifacts. Light curves have been analyzed both with the Lomb-Scargle periodogram 
and with the autocorrelation methods. After a careful check of the consistency among the results 
and a visual inspection of all the original and folded light curves we retain 1978 periods. After 
having identified eclipsing systems and pulsators we identify 1727 rotational periods.

Taking advantage of the more recent version of the gyrochronology relations, we have been able to 
identify more than 500 stars in our sample which are likely younger than 600 Myr.
We cannot use our rotational period distribution to infer the age distribution in the solar neighborhood 
since we cannot assess the completeness of our sample. In fact the original sample observed by CoRoT was 
selected in order to maximize the probability to detect exoplanetary systems eliminating stars affected by crowding. 
The biases introduced by this choice are not obvious. Furthermore the data analysis procedure has 
discarded a non-negligible fraction of light curves due to instrumental artifacts. Finally, as 
discussed in the previous section, we may derive periods only in stars with clean light curves and 
with amplitude variations measurable in our observations.

However our analysis has allowed us to identify pulsators, binaries and, among a huge number of field stars, a sample 
dominated by young stars with small contamination by binary systems.
The nature of our selected sample of young stars is consistent with the nature of the population identified by X-ray 
limited surveys that shown the presence of an excess of young stars in the solar neighborhood.
 
Future spectroscopic follow-up observations are needed. 
This will also allow us to derive chemical abundances, in particular lithium, and
radial velocities, as well as to obtain information from several activity indicators, for instance the 
emission of chromospheric features such as Ca~II H and K. 

Spectroscopic observations will allow us to obtain an independent estimate of the rotation of 
our stars through the measure of the projected rotational velocity, $v {\rm sin}\,i$. This measure may 
be extended to a sample of stars for which we were unable to measure the period in particular for stars with chaotic LCs, in order to better 
understand the biases in our period determinations (i.e. to show if these stars are mainly slow 
rotators).

\section*{Acknowledgments}

The authors would like to thank the referee, Prof. Basri, for his careful reading of the manuscript and for his helpful suggestions to improve this paper. We also thank Dr. J. Bouvier for helpful discussions regarding the
interpretation of the results.\\
L. A. and G. M. acknowledge support from the ASI-INAF agreement I/044/10/0.

\label{lastpage}


\begin{thebibliography}{99}

\bibitem[\protect\citeauthoryear{Affer et al.} {2008}]{amm08} Affer, L., Micela, G., \& Morel, T.\ 2008, A\&A, 483, 801 
\bibitem[\protect\citeauthoryear{Alapini \& Aigrain} {2008}]{aa08} Alapini, A., \& Aigrain, S.\ 2008, arXiv:0801.1237 
\bibitem[\protect\citeauthoryear{Aigrain et al.} {2004}]{aig04} Aigrain, S., \& Irwin, M.\ 2004, MNRAS, 350, 331 
\bibitem[\protect\citeauthoryear{Aigrain et al.} {2009}]{aig09} Aigrain, S., et al.\ 2009, A\&A, 506, 425 
\bibitem[\protect\citeauthoryear{Baglin et al.} {2006}]{baglin06} Baglin, A., et al.\ 2006, 36th COSPAR Scientific Assembly, 36, 3749
\bibitem[\protect\citeauthoryear{Barnes} {2003}]{bar03} Barnes, S.~A.\ 2003, ApJ, 586, 464
\bibitem[\protect\citeauthoryear{Barnes} {2007}]{bar07} Barnes, S.~A.\ 2007, ApJ, 669, 1167  
\bibitem[\protect\citeauthoryear{Basri et al.} {2010}]{bwb+10} Basri, G., Walkowicz, L.~M., Batalha, N., et al.\ 2010, ApJL, 713, L155 
\bibitem[\protect\citeauthoryear{Basri et al.} {2011}]{bwb+11} Basri, G., Walkowicz, L.~M., Batalha, N., et al.\ 2011, ApJ, 141, 20
\bibitem[\protect\citeauthoryear{Bord{\'e} et al.} {2007}]{bfo+07} Bord{\'e}, P., Fressin, F., Ollivier, M., L{\'e}ger, A., \& Rouan, D.\ 2007, Transiting Extrapolar Planets Workshop, 366, 145 
\bibitem[\protect\citeauthoryear{Bouvier et al.} {1997}]{bfa97} Bouvier, J., Forestini, M., \& Allain, S.\ 1997, A\&A, 326, 1023 
\bibitem[\protect\citeauthoryear{Bouvier et al.} {2007}]{bouvier07} Bouvier, J., \& Appenzeller, I.\ 2007, IAU Symposium, 243  
\bibitem[\protect\citeauthoryear{Box \& Jenkins} {1976}]{box76} Box, G.~E.~P., \& Jenkins, G.~M.\ 1976, Holden-Day Series in Time Series Analysis, Revised ed., San Francisco: Holden-Day, 1976 
\bibitem[\protect\citeauthoryear{Cabrera et al.} {2009}]{cfo+09} Cabrera, J., Fridlund, M., Ollivier, M., et al.\ 2009, A\&A, 506, 501 
\bibitem[\protect\citeauthoryear{Carone et al.} {2011}]{cgc+11} Carone, L., Gandolfi, D., Cabrera, J., et al.\ 2011, arXiv:1110.2384 
\bibitem[\protect\citeauthoryear{Carpano \& Fridlund} {2008}]{cf08} Carpano, S., \& Fridlund, M.\ 2008, A\&A, 485, 607 
\bibitem[\protect\citeauthoryear{Collier Cameron et al.} {2009}]{ccd09} Collier Cameron, A., et al.\ 2009, MNRAS, 400, 451 
\bibitem[\protect\citeauthoryear{Cutri et al.} {2003}]{csv+03} Cutri, R.~M., Skrutskie, M.~F., van Dyk, S., et al.\ 2003, ``The IRSA 2MASS All-Sky Point Source Catalog, NASA/IPAC Infrared Science Archive.'' 
\bibitem[\protect\citeauthoryear{Dahm \& Simon} {2005}]{ds05} Dahm, S.~E., \& Simon, T.\ 2005, AJ, 129, 829 
\bibitem[\protect\citeauthoryear{Debosscher et al.} {2009}]{dsl+09} Debosscher, J., Sarro, L.~M., L{\'o}pez, M., et al.\ 2009, A\&A, 506, 519 
\bibitem[\protect\citeauthoryear{Della Ceca et al.} {2004}]{dmc+04} Della Ceca, R., et al.\ 2004, A\&A, 428, 383 
\bibitem[\protect\citeauthoryear{Deleuil et al.} {2006}]{del06} Deleuil, M., Moutou, C., Deeg, H.~J., et al.\ 2006, ESA Special Publication, 1306, 341
\bibitem[\protect\citeauthoryear{Deleuil et al.} {2009}]{del09} Deleuil, M., et al.\ 2009, AJ, 138, 649 
\bibitem[\protect\citeauthoryear{Duquennoy \& Mayor} {1991}]{dm91} Duquennoy, A., \& Mayor, M.\ 1991, A\&A, 248, 485 
\bibitem[\protect\citeauthoryear{Eaton et al.} {1995}]{ehh95} Eaton, N.~L., Herbst, W., \& Hillenbrand, L.~A.\ 1995, AJ, 110, 1735 
\bibitem[\protect\citeauthoryear{Favata et al.} {1988}]{fsr+88} Favata, F., Sciortino, S., Rosner, R., \& Vaiana, G.~S.\ 1988, ApJ, 324, 1010 
\bibitem[\protect\citeauthoryear{Favata et al.} {1993}]{fbm+93} Favata, F., Barbera, M., Micela, G., \& Sciortino, S.\ 1993, A\&A, 277, 428 
\bibitem[\protect\citeauthoryear{Flaccomio et al.} {2005}]{flaccomio05} Flaccomio, E., Micela, G., Sciortino, S., Feigelson, E.~D., Herbst, W., Favata, F., Harnden, F.~R., Jr., \& Vrtilek, S.~D.\ 2005, ApJS, 160, 450 
\bibitem[\protect\citeauthoryear{Gioia et al.} {1990}]{gms+90} Gioia, I.~M., Maccacaro, T., Schild, R.~E., Wolter, A., Stocke, J.~T., Morris, S.~L., \& Henry, J.~P.\ 1990, ApJ, 72, 567 
\bibitem[\protect\citeauthoryear{Henry et al.} {2001}]{hgm+01} Henry, J.~P., Gioia, I.~M., Mullis, C.~R., Voges, W., Briel, U.~G., B{\"o}hringer, H., \& Huchra, J.~P.\ 2001, ApJ, 553, L109 
\bibitem[\protect\citeauthoryear{Horne \& Baliunas} {1986}]{hb86} Horne, J.~H., \& Baliunas, S.~L.\ 1986, ApJ, 302, 757 
\bibitem[\protect\citeauthoryear{Irwin et al.} {2006}]{iah+06} Irwin, J., Aigrain, S., Hodgkin, S., Irwin, M., Bouvier, J., Clarke, C., Hebb, L., \& Moraux, E.\ 2006, MNRAS, 370, 954 
\bibitem[\protect\citeauthoryear{Irwin et al.} {2011}]{ibb+11} Irwin, J., Berta, Z.~K., Burke, C.~J., et al.\ 2011, ApJ, 727, 56 
\bibitem[\protect\citeauthoryear{Kendall et al.} {2005}]{kbm+05} Kendall, T.~R., Bouvier, J., Moraux, E., James, D.~J., \& M{\'e}nard, F.\ 2005, A\&A, 434, 939 
\bibitem[\protect\citeauthoryear{Kraft} {1967}]{kra67} Kraft, R.~P.\ 1967, ApJ, 150, 551 
\bibitem[\protect\citeauthoryear{Lamm et al.} {2004}]{lbm+04} Lamm, M.~H., Bailer-Jones, C.~A.~L., Mundt, R., Herbst, W., \& Scholz, A.\ 2004, A\&A, 417, 557 
\bibitem[\protect\citeauthoryear{Lamm et al.} {2005}]{lmb+05} Lamm, M.~H., Mundt, R., Bailer-Jones, C.~A.~L., \& Herbst, W.\ 2005, A\&A, 430, 1005 
\bibitem[\protect\citeauthoryear{Lomb} {1976}]{lom76} Lomb, N.~R.\ 1976, Ap\&SS, 39, 447
\bibitem[\protect\citeauthoryear{L{\'o}pez-Santiago et al.} {2006}]{lmc+06} L{\'o}pez-Santiago, J., Montes, D., Crespo-Chac{\'o}n, I., \&  Fern{\'a}ndez-Figueroa, M.~J.\ 2006, ApJ, 643, 1160
\bibitem[\protect\citeauthoryear{L{\'o}pez-Santiago et al.} {2007}]{lms+07} L{\'o}pez-Santiago, J., Micela, G., Sciortino, S., Favata, F., Caccianiga, A., Della Ceca, R., Severgnini, P., \& Braito, V.\ 2007, A\&A, 463, 165 
\bibitem[\protect\citeauthoryear{Mamajek \& Hillenbrand} {2008}]{mh08} Mamajek, E.~E., \& Hillenbrand, L.~A.\ 2008, ApJ, 687, 1264 
\bibitem[\protect\citeauthoryear{Mayor et al.} {1992}]{mdh+92} Mayor, M., Duquennoy, A., Halbwachs, J.-L., \& Mermilliod, J.-C.\ 1992, IAU Colloq.~135: Complementary Approaches to Double and Multiple Star Research, 32, 73 
\bibitem[\protect\citeauthoryear{Meibom et al.} {2009}]{mms09} Meibom, S., Mathieu, R.~D., \& Stassun, K.~G.\ 2009, ApJ, 695, 679 
\bibitem[\protect\citeauthoryear{Messina et al.} {2008}]{mdp+08} Messina, S., Distefano, E., Parihar, P., Kang, Y.~B., Kim, S.-L., Rey, S.-C., \& Lee, C.-U.\ 2008, A\&A, 483, 253 
\bibitem[\protect\citeauthoryear{Meunier et al.} {2007}]{mdm+07} Meunier, J.-C., Deleuil, M., Moutou, C., Ouchani, M., Savalle, R., \& Surace, C.\ 2007, Astronomical Data Analysis Software and Systems XVI, 376, 339 
\bibitem[\protect\citeauthoryear{Micela et al.} {2007}]{maf+07} Micela, G., Affer, L., Favata, F., Henry, J.~P., Gioia, I., Mullis, C.~R., Sanz Forcada, J., \& Sciortino, S.\ 2007, A\&A, 461, 977 
\bibitem[\protect\citeauthoryear{Radick et al.} {1987}]{rtl+87} Radick, R.~R., Thompson, D.~T., Lockwood, G.~W., Duncan, D.~K., \& Baggett, W.~E.\ 1987, ApJ, 321, 459 
\bibitem[\protect\citeauthoryear{Rebull et al.} {2002}]{rms+02} Rebull, L.~M., et al.\ 2002, AJ, 123, 1528 
\bibitem[\protect\citeauthoryear{Renner et al.} {2008}]{rre+08} Renner, S., Rauer, H., Erikson, A., et al.\ 2008, A\&A, 492, 617 
\bibitem[\protect\citeauthoryear{Robin \& Creze} {1986}]{rc86} Robin, A., \& Creze, M.\ 1986, A\&A, 157, 71 
\bibitem[\protect\citeauthoryear{Robin et al.} {2003}]{rrd+03} Robin, A.~C., Reyl{\'e}, C., Derri{\`e}re, S., \& Picaud, S.\ 2003, A\&A, 409, 523 
\bibitem[\protect\citeauthoryear{Ruphy et al.} {1997}]{rec+97} Ruphy, S., Epchtein, N., Cohen, M., et al.\ 1997, A\&A, 326, 597 
\bibitem[\protect\citeauthoryear{Samadi et al.} {2007}]{sfc+07} Samadi, R., Fialho, F., Costa, J.~E.~S., Drummond, R., Pinheiro Da Silva, L., Baudin, F., Boumier, P., \& Jorda, L.\ 2007, arXiv:astro-ph/0703354
\bibitem[\protect\citeauthoryear{Scargle} {1982}]{sca82} Scargle, J.~D.\ 1982, ApJ, 263, 835
\bibitem[\protect\citeauthoryear{Schatzman} {1962}]{sch62} Schatzman, E.\ 1962, Annales d'Astrophysique, 25, 18 
\bibitem[\protect\citeauthoryear{Sciortino et al.} {1995}]{sfm95} Sciortino, S., Favata, F., \& Micela, G.\ 1995, A\&A, 296, 370 
\bibitem[\protect\citeauthoryear{Schmidt et al.} {2004}]{sjl+04} Schmidt, E.~G., Johnston, D., Langan, S., \& Lee, K.~M.\ 2004, ApJ, 128, 1748
\bibitem[\protect\citeauthoryear{Schmidt et al.} {2005a}]{sjl+05a} Schmidt, E.~G., Johnston, D., Langan, S., \& Lee, K.~M.\ 2005, ApJ, 129, 2007 
\bibitem[\protect\citeauthoryear{Schmidt et al.} {2005b}]{sjl+05b} Schmidt, E.~G., Johnston, D., Langan, S., \& Lee, K.~M.\ 2005, ApJ, 130, 832 
\bibitem[\protect\citeauthoryear{Skumanich} {1972}]{sku72} Skumanich, A.\ 1972, ApJ, 171, 565 
\bibitem[\protect\citeauthoryear{Vrba et al.} {1988}]{vrba88} Vrba, F.~J., Herbst, W., \& Booth, J.~F.\ 1988, ApJ, 96, 1032  
\bibitem[\protect\citeauthoryear{Weldrake \& Bayliss} {2008}]{wb08} Weldrake, D.~T.~F., \& Bayliss, D.~D.~R.\ 2008, ApJ, 135, 649 
\bibitem[\protect\citeauthoryear{Wilson} {1966}]{wil66} Wilson, O.~C.\ 1966, ApJ, 144, 695 
\bibitem[\protect\citeauthoryear{Zuckerman \&  Song} {2004}]{zs04} Zuckerman, B., \&  Song, I.\ 2004, ARAA, 42, 685

\end{thebibliography}
\end{document}